\documentclass[preprint]{aastex61}
\usepackage{amssymb,amsmath}
\usepackage{graphicx,epsfig}

\begin{document}

\title{A study of a tilted thin inner accretion disk around a spinning black hole}

\correspondingauthor{Srimanta Banerjee}
\email{srimanta.banerjee@tifr.res.in}

\author{Srimanta Banerjee}
\affiliation{Department of Astronomy and Astrophysics, Tata Institute of Fundamental Research, Mumbai 400005, India}

\author{Chandrachur Chakraborty}
\affiliation{Kavli Institute for Astronomy and Astrophysics, Peking University, Beijing 100871, China}

\author{Sudip Bhattacharyya}
\affiliation{Department of Astronomy and Astrophysics, Tata Institute of Fundamental Research, Mumbai 400005, India}

\begin{abstract}
The inner part of a thin accretion disk around a Kerr black hole can serve as an important tool to study 
the physics of the strong gravity regime. A tilt in such a disk with respect to the black hole spin 
axis is particularly useful for this purpose, as such a tilt can have a significant effect on the 
observed X-ray spectral and timing features via Lense-Thirring precession. However, the inner disk 
has been predicted to become aligned with the spin direction of the black hole by the well-known 
Bardeen-Petterson effect. 
Here we calculate, both analytically and numerically, the radial profile of the thin accretion disk 
tilt angle in the viscous regime (i.e., $\alpha > H/R$; $\alpha$ is the Shakura-Sunyaev viscosity parameter,
$H$ is the disk thickness and $R$ is the radial distance).
We show that the inner disk may not be aligned at all for certain reasonable ranges of parameter values.
This makes the inner accretion disk particularly promising to probe the black hole parameters, and the accretion process in the strong gravity region.   
\end{abstract}

\keywords{accretion, accretion disks --- black hole physics --- diffusion --- relativistic processes --- X-rays: binaries.}

\section{Introduction}\label{intro}
Warped accretion disks, i.e., disks with planes changing with the radius, 
are found in a wide variety of systems, e.g., protostars, X-ray binaries and active galactic nuclei (AGN). 
For example, warped disks are observed in the AGN NGC 4258 \citep{Herrnstein,Papaloizou}, the X-ray binaries SS433 \citep{Begelman}, Her X-1 \citep{Wijers1999} and GRO J1655-40 \citep{Martin}, and pre-main-sequence stars KH 15D \citep{Chiang,Facchini} and HD 142527 \citep{Cas}. Although these warps are ubiquitous in astrophysics, the torques producing such a warp can have different origins, and depend on the astrophysical system under consideration.
Usually, if there is a non-axisymmetric force acting on such a disk, a warp is generated. 
An initially planar disk may become warped because of relativistic effects induced by a misaligned black hole \citep{BP75}, radiation effect \citep{Pringle96}, or tidal interaction with a companion star \citep[for protostellar disks; ][]{Larwood}.

An analytical theory for such a warped thin disk around a spinning black hole was developed by Bardeen and Petterson in 1975 \citep{BP75}. They proposed that the relativistic Lense-Thirring precession (LT) \citep{LT} may have a significant effect on the disk around a misaligned black hole, and aligns the inner disk with the spin direction of the black hole. This is the Bardeen-Petterson (BP) effect. As the LT torque falls off with the radial distance ($R$) rapidly (roughly as $1/R^3$), the viscous torque gradually takes over. Hence the outer part remains tilted with respect to the spin axis of the black hole, whereas the inner part of the disk stays aligned. 
In between these two regions, the disk tilt angle gradually transits from an aligned inner disk to a 
misaligned outer disk, resulting in a warp in the disk. 
Although the transfer of the angular momentum in a warped disk is three dimensional, 
the above mentioned work, along with  \cite{Petterson77a}, \cite{Petterson77b} and \cite{Hatchett} where similar scenarios were investigated, considered viscosity to be described by the same turbulent viscosity $\nu=\alpha c_s H$ \citep{Shakura}, where $c_s$ is the local sound speed, $H$ is the thickness of the disk and $\alpha$ is the Shakura-Sunyaev parameter, in all directions. Besides, the evolution equations that they had considered, did not conserve the angular momentum. The first self-consistent analysis of a thin warped viscous disk was possibly done by \cite{Pap83}, where they considered two viscosities arising due to azimuthal and vertical shears, 
and obtained the evolution equations of the warped disk in two ways. In their `naive' approach, they generalized the standard treatment of the flat disk, and took into account the correct forms of the torques corresponding to two viscosities to conserve the angular momentum on each annulus of the disk. They also considered detailed analyses of internal fluid dynamics in linear regime, and obtained evolution equation for a slightly tilted (tilt angle $\ll$ $H/R$) viscous disk ($\alpha>H/R$). Later, \cite{Pringle92}, following the `naive' approach of \cite{Pap83}, obtained the general evolution equation of a warped viscous disk, valid for arbitrary warps and viscosities, simply by demanding mass and angular momentum to be conserved in each annulus of the disk. In \cite{Pringle92}, the BP effect was also considered, and numerical simulations were done to study the evolution of the disk tilt angle radial profile in the presence of the LT precession. 
The linear hydrodynamical analysis of \cite{Pap83} was extended to the non-linear regime later by \cite{Ogilvie1999}, and the evolution equations, valid for arbitrary warp amplitudes, were obtained. This derivation confirmed the equations obtained by \cite{Pringle92} in the viscous regime, except for two differences, which can be neglected for small amplitude warps, and small $\alpha$. \cite{SF96} examined the Pringle's equation with the LT torque in steady state limit, solved the differential equation to obtain the radial profile of the disk tilt angle assuming viscosities to be independent of radius, and used this radial profile to calculate the alignment timescale for the black hole. But the contribution of the inner disk was neglected in the \cite{SF96}, as they assumed a priori that the inner disk is aligned up to a radius with the spin direction of the black hole. \cite{Armitage} considered the viscosities to be functionally dependent upon radius, matched their numerical result with \cite{SF96}, and also calculated the alignment time scale for supermassive black holes in AGN with a realistic profile of column density. Later, \cite{Rebecca} also assumed a power law form of viscosities and surface density with the same power law factor for all of them, obtained the steady state profile for the disk tilt angle analytically, and calculated alignment timescale as well as precession timescale for black hole for different values of the power law factor. \cite{Chen} improved the above analytical calculations by considering two viscosities to be a function of $R$ with different power law factors, and obtained an analytical solution of the warped disk equation, apart from giving the numerical solution. 
But all of them assumed the alignment of the inner disk a priori, and therefore mainly focused on the behavior of the  warped region of the disk. 

Interestingly, \cite{Lodato} numerically solved the time-dependent evolution equation of a warped viscous disk obtained by \cite{Pringle92}, and compared their disk inclination radial profile with the steady state radial profiles obtained by \cite{SF96}. \cite{Lodato} found that, for certain ranges of parameter values, their solution deviated largely from what \cite{SF96} had obtained. They also showed that there are scenarios where the inner disk may not be able to align itself with the black hole spin direction, contrary to what has been assumed by some authors. However, the misalignment of the inner accretion disk was earlier shown by \cite{Lubow} in the context of a thick disk or a low viscous disk, in which the warping disturbances are transmitted through waves (not diffusively as in the case of a thin viscous disk). Recently, \cite{Zhuravlev} found no evidence for alignment of the inner disk in GRMHD simulations for their semi-analytic models of moderately thin prograde tilted accretion disks, but found partial alignment for retrograde disks. On the observational side, X-ray observations of the black hole H1743-322 have suggested that the inner disk could be tilted \citep{Ingram, Tom}. 
\citet[][hereafter CB17]{CB17} solved the full steady state warped disk equation analytically for the first time, using the \cite{SF96} formalism and considering the contribution of the inner disk, and obtained an analytical expression for the tilt angle up to first order in Kerr parameter. 

Here, we revisit the case of a tilted thin accretion disk around a Kerr black hole in the viscous regime motivated by the observations of \citep{Ingram, Tom} and the theoretical investigations of \cite{Lodato}, \cite{Zhuravlev} and CB17. In this work, we develop a model of the warped disk following the formalism of \cite{Pringle92}, and study the steady state behavior of the entire disk as a function of several parameters like the Kerr parameter, inner edge disk tilt, etc., 
which can be useful to confront observations.
We take into account the contributions of the inner disk, which were ignored in many earlier works, in our calculations. 
We solve the full warped disk equation (equation (6) of \citep{SF96}) analytically, show explicitly the incompleteness of the analytical expression obtained in CB17, and obtain the correct expression for disk tilt angle up to first order in the Kerr parameter. We numerically solve the full warped disk equation in the steady state to obtain the radial profile of the disk tilt angle, and explore how the radial profile of the tilt angle 
depends on different sets of parameter values. Also, as our solution is valid for the entire disk, we probe the behavior of the inner disk explicitly as a function of the parameters of the system.

This paper is organized as follows. In section \ref{formalism}, we derive all the equations required for the analyses roughly following the discussions in \cite{Pringle92} and \cite{SF96}. We solve the warped disk equations analytically to obtain an expression for tilt angle up to the first order in the Kerr parameter in section \ref{an}, and discuss the numerical setup for solving the same equations in section \ref{num}. 
We discuss the implications of our inner boundary conditions on viscous torques in section \ref{vis}.
The results obtained from analytical as well as numerical computations, and a brief description of 
their implications are given in section \ref{result}. We summarize our work in section \ref{sum}.

\section{Formalism}\label{formalism}
We consider a geometrically thin (i.e., the disk aspect ratio $H/R\ll1$), and a Keplerian accretion disk around a slowly rotating Kerr black hole at the center. The Kerr black hole is described by its mass $M$, and the Kerr parameter $a$. The spin axis of the black hole is directed along the $z$ axis in our coordinate system. The disk is tilted with respect to the spin axis of the black hole.

We divide the disk into circular rings of width $\Delta R$, and define surface density $\Sigma(R,t)$, radial velocity $V_R(R,t)$ and angular momentum density (angular momentum per unit surface area of the disk) $\textbf{L}(R,t)=\Sigma R^2 \Omega(R)\textbf{l}(R,t)$ on each annulus of the disk, where $\textbf{l}$ is the unit tilt vector directed normal to the plane of the disk and $\Omega(R)$ is the Keplerian angular speed. We assume the tilt angle to be small, i.e. $\textbf{l}\simeq(l_{x},l_{y},1)$, and the disk to be sufficiently viscous, i.e., $\alpha>H/R$. In this viscous regime, warp is transported diffusively in the disk \citep{Pap83}. However, in the opposite regime, i.e. $\alpha<H/R$ , which would not be considered in this work, warping disturbances propagate in a wave-like manner \citep{Lubow, Ivanov}.

The local mass, and angular momentum conservation equations in the viscous regime ($\alpha>H/R$) take the following form \citep{Pap83,Pringle92}
\begin{equation}\label{MD}
\frac{\partial \Sigma}{\partial t}+\frac{1}{R}\frac{\partial}{\partial R}\left(RV_R\Sigma\right)=0,
\end{equation}
and 
\begin{equation}\label{AM2}
\frac{\partial }{\partial t}\left(\Sigma R^{2}\Omega \textbf{l}\right)+\frac{1}{R}\frac{\partial}{\partial R}\left(\Sigma V_R R^3 \Omega \textbf{l}\right)=\frac{1}{R}\frac{\partial}{\partial R}\left(\nu_1 \Sigma R^3 \Omega^{'}\textbf{l}\right)+\frac{1}{R}\frac{\partial}{\partial R}\left(\frac{1}{2}\nu_2 \Sigma R^3 \Omega \frac{\partial \textbf{l}}{\partial R}\right),
\end{equation}
where $\nu_1$  is the viscosity associated with the azimuthal shear, i.e., $(R,\phi)$ component of shear, $\nu_2$ is  the viscosity associated with the vertical shear, i.e., $(R,z)$ component of shear \citep{Pap83}, and $\Omega^{'}=d\Omega/dR$. The first torque term on the right hand side (rhs), which acts perpendicular to the plane of the disk, appears also in the case of a flat disk (i.e., $\partial\textbf{l}/\partial R$ is zero). On the other hand, the second torque term acting in the plane of the disk arises, only when the disk is warped. The first torque term acts on the differential rotation in the plane of the disk driving the accretion process, where the second term helps to make the disk flat. The ratio between the two viscosities, i.e. viscous anisotropy $\nu_2/\nu_1$, can be shown to be related to $\alpha$ for a small amplitude warp in the following way \citep{Ogilvie1999}
\begin{equation}\label{alpha}
\frac{\nu_2}{\nu_1}=\frac{1}{2\alpha^2}.\frac{4(1+7\alpha^2)}{4+\alpha^2}.
\end{equation} 
In our formalism, we assume both these viscosities $\nu_1$ and $\nu_2$ to be constant. 

The expression for radial velocity can be obtained from the above Equations (\ref{MD}) and  (\ref{AM2}). It gives \citep{Pringle92}
\begin{equation}\label{VR}
V_R=\frac{\partial/\partial R\left(\nu_1\Sigma R^3 \Omega^{'}\right)-\frac{1}{2}\nu_2\Sigma R^3 \Omega\vert\partial \textbf{l}/\partial R\vert^2}{R\Sigma(\partial/\partial R)\left(R^2 \Omega\right)}.
\end{equation}
Since we are considering the tilt angle to be small in this paper, the term $\vert\partial \textbf{l}/\partial R\vert^2$ can be dropped. 

If the expression for radial velocity (equation (\ref{VR})) is used to substitute the same in the local angular momentum density Equation (\ref{AM2}), one gets \citep{Pringle92}
\begin{equation}\label{AM1}
\frac{\partial \textbf{L}}{\partial t}=\frac{1}{R}\frac{\partial}{\partial R}\left[\frac{\frac{\partial}{\partial R}\left\lbrace\nu_1\Sigma R^3\left(-\Omega^{'}\right)\right\rbrace}{\Sigma\frac{\partial}{\partial R}\left(R^2\Omega\right)}\textbf{L}\right]+\frac{1}{R}\frac{\partial}{\partial R}\left[\frac{1}{2}\nu_2 R \vert\textbf{L}\vert \frac{\partial \textbf{l}}{\partial R}\right]+\frac{1}{R}\frac{\partial}{\partial R}\left[\nu_1 \left(\frac{R\Omega^{'}}{\Omega}\right)\textbf{L}\right].
\end{equation}
The above equation describes the evolution of angular momentum density for an annulus of the thin tilted disk in the diffusive regime. The first two terms on the right hand side of the above equation are diffusive, whereas the last term is advective \citep{Pringle92}. 
Here, we stress that, although the equation (\ref{AM1}) does not emerge from the discussions of the internal hydrodynamics of the system, \cite{Ogilvie1999} has shown that the above equation is valid for small tilt angles and small $\alpha$ (see sections 7.1 and 7.3 of \cite{Ogilvie1999}) from the detailed three dimensional hydrodynamical analysis of warped viscous disk in the non-linear regime. Later, \cite{Price2010} have found a remarkable agreement between this analytic theory, and their numerical investigations of a warped accretion disk in viscous regime from SPH (smoothed particle hydrodynamic) simulations.

In the case of a Keplerian thin disk around a slowly rotating black hole \citep{Pringle92}, where
\begin{equation}\label{omega}
\Omega(R)=\sqrt{GM}R^{-3/2},
\end{equation}
and
\begin{equation}\label{L}
\textbf{L}(R,t)=\sqrt{GMR}\Sigma \textbf{l}(R,t),
\end{equation}
the above evolution equation becomes
\begin{equation}\label{AM}
\frac{\partial \textbf{L}}{\partial t}=\frac{1}{R}\frac{\partial}{\partial R}\left[\frac{3R^{1/2}}{\Sigma}\frac{\partial}{\partial R}\left(\nu_1 \Sigma R^{1/2}\right)\textbf{L}-\frac{3}{2}\nu_1 \textbf{L}+\frac{1}{2}\nu_2 R \vert \textbf{L}\vert \frac{\partial \textbf{l}}{\partial R}\right].
\end{equation}

Equation (\ref{AM}) describes the evolution of angular momentum density subjected to internal torques only. In order to capture the relativistic effect acting on the disk as a consequence of the presence of a Kerr black hole, we will have to add the external torque due to LT precession, given by
\begin{equation}\label{LT}
\frac{\partial \textbf{L}}{\partial t}=\boldsymbol{\Omega}_p\times\textbf{L},
\end{equation}
to the rhs of the Equation (\ref{AM}) \citep{Pringle92}. The precession rate $\Omega_p$ is given by $\Omega_p\approx\omega_p/R^3$ in the slow rotation limit of Kerr black hole (CB17). Here,
\begin{equation}
\omega_p=\frac{2GJ_{\rm BH}}{c^2}=\frac{2a G^2 M^2}{c^3}=2acR_g^2,
\end{equation}
where $G$ is the Newtonian gravitational constant, $c$ is the speed of light in free space, $J_{\rm BH}$ is the angular momentum of the black hole, and $R_g=GM/c^2$ is the gravitational radius. 

In this paper, we are interested in studying the interplay between the viscous and LT torques in the disk in steady state, and how their relative dominance determines the essential physics of the disk. In steady state, the evolution Equation (\ref{AM}) subjected to LT torque takes the following form \citep{SF96}
\begin{equation}\label{M}
\frac{1}{R}\frac{\partial}{\partial R}\left[\left(\frac{3R}{L}\frac{\partial}{\partial R}\left(\nu_1 L\right)-\frac{3}{2}\nu_1 \right)\textbf{L}+\frac{1}{2}\nu_2 R L \frac{\partial \textbf{l}}{\partial R}\right]+\frac{\boldsymbol{\omega}_p\times\textbf{L}}{R^3}=0,
\end{equation}
where we have used the expression of $L=\vert\textbf{L}\vert$ (Equation (\ref{L})). 

The distribution of $L(R)$ in the disk in steady state can be obtained by taking the scalar product of $\textbf{l}$ with the Equation ($\ref{M}$). It gives \citep{SF96}
\begin{equation}
\frac{1}{R}\frac{\partial}{\partial R}\left[R\frac{\partial}{\partial R}\left(\nu_1 L\right)-\frac{1}{2}\nu_1 L\right]=0,
\end{equation}
under the small tilt angle approximation (we have ignored the term $\vert\partial \textbf{l}/\partial R\vert^2$). Solving the above equation, one obtains 
\begin{equation}\label{Ld}
L(R)=C_2 R^{1/2}-2 C_1,
\end{equation}
where $C_1$ and $C_2$ are the integration constants. Upon using the Equations (\ref{L}) and (\ref{Ld}), one obtains
\begin{equation}
C_2=\sqrt{GM}\Sigma+2 C_1 R^{-1/2}.
\end{equation}
Now the expression for $C_2$ can be derived from the above equation upon using the boundary condition $\Sigma \rightarrow \Sigma_{\infty}$ as $R \rightarrow \infty$. It gives \citep{SF96}
\begin{equation}\label{C2}
C_2=\sqrt{GM}\Sigma_{\infty}.
\end{equation}
In order to find an expression for $C_1$, we substitute the expression of $C_2$ (Equation (\ref{C2})) into Equation (\ref{Ld}), use the inner edge boundary condition $\Sigma(R_{\rm in})=\Sigma_{\rm in}$ 
\citep[$\Sigma_{\rm in}>\Sigma_{\infty}$; ][CB17]{Lubow}, and we get
\begin{equation}\label{C1}
C_1=\frac{1}{2}\sqrt{GMR_{\rm in}}\left(\Sigma_{\infty}-\Sigma_{\rm in}\right),
\end{equation}
where $R_{\rm in}$ corresponds to the inner edge radius of the disk, which is essentially identical to the ISCO radius $R_{\rm ISCO}$ for a Kerr black hole. Upon substituting the above expressions for $C_1$ and $C_2$ into Equation (\ref{Ld}), we obtain the following expression for $L(R)$ in steady state: 
\begin{equation}\label{gl}
L(R)=\sqrt{GM}\left[R^{1/2}\Sigma_{\infty}+R^{1/2}_{\rm in}\left(\Sigma_{\rm in}-\Sigma_{\infty}\right)\right].
\end{equation}
Since the inner disk was always assumed to be aligned with the black hole spin axis, the behavior of inner disk was not mostly considered earlier. Consequently, as the term $C_1$ is associated with the inner edge boundary condition, $C_1$ was mostly ignored in the steady state expression of L (equation \ref{gl}). 

The steady state distribution of the surface density can similarly be obtained from the above equation (\ref{gl}) upon substituting the rhs by Equation (\ref{L})
\begin{equation}\label{sig}
\Sigma(R)=\Sigma_{\infty}+\left(R_{\rm in}/R\right)^{\frac{1}{2}}\left(\Sigma_{\rm in}-\Sigma_{\infty}\right).
\end{equation}
Although the contribution of $C_1$ was mostly neglected earlier, several different functional forms of surface density have been considered. For example, \cite{Rebecca} assumed $\Sigma(R)$ of the form $R^{-\beta}$ where $\beta$ is a free parameter. Therefore, from Equations (\ref{gl}) and (\ref{sig}), we find that in steady state $L(R)\propto R^{1/2}$ and $\Sigma\propto R^{-1/2}$ as a consequence of the diffusion process driving between the two edges of the disk \citep{Frank}.

Now substituting equation (\ref{Ld}) into (\ref{M}), we obtain the following equations for $l_x$ and $l_y$ (i.e., $x$ and $y$ components of tilt vector) in steady state
\begin{equation}\label{warp1}
\frac{\partial}{\partial R}\left(3\nu_1 C_1 l_x+\frac{1}{2}\nu_2 R L \frac{\partial l_x}{\partial R}\right)=\omega_p \frac{L}{R^2}l_y,
\end{equation} 
and 
\begin{equation}\label{warp2}
\frac{\partial}{\partial R}\left(3\nu_1 C_1 l_y+\frac{1}{2}\nu_2 R L \frac{\partial l_y}{\partial R}\right)=-\omega_p \frac{L}{R^2}l_x,
\end{equation} 
where  $\boldsymbol{\omega}_p\times\textbf{l}=\left(-\omega_p l_y,\omega_p l_x,0\right)$ in our construction.  We can also combine the above equations to arrive at \citep{SF96}
\begin{equation}\label{warp}
\frac{\partial}{\partial R}\left(3\nu_1 C_1 W+\frac{1}{2}\nu_2 R L \frac{\partial W}{\partial R}\right)=-i\omega_p \frac{L}{R^2}W,
\end{equation} 
where $W=l_x+i l_y=\beta e^{i\gamma}$. Here, $\beta=\sqrt{l_x^2+l_y^2}$ and $\gamma=\tan^{-1}\left(l_y/l_x\right)$ represent the tilt angle and twist angle respectively.  The Equation (\ref{warp}) or Equations (\ref{warp1}) and (\ref{warp2})  (we would be referring them as warped disk equation(s) hereafter) encapsulate the basic features of a warped accretion disk around a spinning black hole in steady state. As the inner disk was always a priori assumed to be aligned with the black hole spin direction due to the BP effect in earlier works, the main focus was paid mostly on the physics of warped part of the disk. As a result, the terms associated with $C_1$ have been ignored in the past (except in CB17, and \cite{Lodato}). Our aim is to solve the full warped disk equation (\ref{warp}) analytically as well as numerically with realistic boundary conditions to obtain the radial profile of the disk tilt angle, and to explore how the radial profile of the disk tilt angle in the inner as well as the outer disk depend on parameters like the Kerr parameter, viscosity etc.
\section{Solution of the warped disk equation} 
\subsection{Analytical Solution of the warped disk equation}\label{an}
In this section we present the analytical solution of the warped disk Equation (\ref{warp}) or Equations (\ref{warp1}) and (\ref{warp2}). We solve the equation(s) using perturbative method, and obtain an expression for the tilt angle $\beta=\vert W \vert$ up to first order in $a$. In order to achieve this, we consider an expansion of $W$ in orders of $a$, 
\begin{equation}
W=l_x^{(0)}+i l_y^{(0)}+a\left(l_x^{(1)}+i l_y^{(1)}\right)+a^2 \left( l_x^{(2)}+i l_y^{(2)}\right)+...,
\end{equation}
where $l_x^{(0)}$, $l_x^{(1)}$, $l_x^{(2)}$ and $l_y^{(0)}$, $l_y^{(1)}$, $l_y^{(2)}$ are the the zeroth, first, and second order terms of the real and complex part of $W$, respectively (see the definition after Equation (\ref{warp})). 
Here, we consider only spinning black holes, as
there is no preferred axis of a non-spinning black hole to define the tilt angle.
Besides, we do not define $l_x^{(0)}$ and $l_y^{(0)}$ for exactly $a = 0$, but we define them 
for $a \approx 0$ (that is, $a$ is extremely small, but nonzero), which is a practical way 
to have a well-defined black hole spin axis to measure the tilt angle. 
We ignore the contribution of $l_y^{(0)}$ as the twist angle becomes negligibly small in the limit of very small LT precession (see the expression of twist angle mentioned just after equation (\ref{warp})). Hence, the expression for tilt angle $\beta$ up to first order in $a$ is given by
\begin{equation}\label{beta_1}
\beta^{(1)}=\sqrt{\left(l_x^{(0)}\right)^2+2 a l_x^{(0)} l_x^{(1)}+2 a^2 l_x^{(0)} l_x^{(2)}+a^2 \left(l_x^{(1)}\right)^2+a^2\left(l_y^{(1)}\right)^2}.
\end{equation}
Therefore, we will have to solve the warped disk Equation (\ref{warp}) up to second order in $a$ in order to obtain the analytical expression of tilt angle $\beta$ up to first order in $a$ (see the equation above, which includes the second order term $l_x^{(2)}$). The expressions of $l_x^{(0)}$, $l_x^{(1)}$ and $l_y^{(1)}$ were already calculated in CB17 by solving the Equation (\ref{warp}) using some specific boundary conditions. In CB17, $l_x^{(0)}$ and $al_y^{(1)}$ were designated by $W_0$ (zeroth order term in the expansion of $W$) and $B$ (imaginary part of $aW_a$, where $W_a$ is the first order term in the expansion of $W$) respectively, and $l_x^{(1)}$ is essentially the term containing $W_{a,\rm in}$, i.e., the real part of $W_{a}$ (see Equations (33) and (35) of CB17; their $A$ has a contribution from both $W_0$ and the real part of $W_{a}$). But in CB17, the authors did not calculate $l_x^{(2)}$ as they did not include the term $2 a^2 l_x^{(0)} l_x^{(2)}$ in their expression of the tilt angle up to first order in $a$ (see equations (35), (36) and (37) of CB17). Hence, the expression of $\beta$ ($=\sqrt{A^2+B^2}$ in their paper) that they got, claimed to be up to first order in $a$, is incomplete. Here, we include all the terms (i.e., all the terms present in Equation (\ref{beta_1})) required to compute the tilt angle up to the first order in $a$, and use general boundary conditions to solve the warped disk Equation (\ref{warp}). Hence, we report here the new corrected analytical expression of the disk tilt angle up to first order in the Kerr parameter. We also explore the importance of the term  $2 a^2 l_x^{(0)} l_x^{(2)}$, which CB17 did not include, and show the hump feature CB17 obtained is an artifact of using an incomplete expression in the section \ref{res1}.

\subsubsection{Analytical solution of the warped disk equation up to first order}\label{old}

Here, we briefly mention the solutions of the Equation (\ref{warp}) up to the first order, as discussed in CB17, for the sake of completeness. Then, in the section \ref{next}, we present the detailed new calculations required for obtaining the second order term $l_x^{(2)}$ (in order to get the expression for $2 a^2 l_x^{(0)} l_x^{(2)}$), which was not considered in CB17. 

The solution of the Equation (\ref{warp1}) of order zero in $a$ is given by (Equation (22) of CB17)
\begin{equation}
l_x^{(0)}=\frac{z^n\left(W_{\infty}-W_{0,\rm in}\right)+\left(W_{\rm 0,in}-W_{\infty} z^n_{\rm in}\right)}{1-z^n_{\rm in}},
\end{equation}
where 
\begin{equation}
z_{\rm in}=1+\frac{2 C_1}{L(R_{\rm in})}=\frac{\Sigma_{\infty}}{\Sigma_{\rm in}},
\end{equation}
and 
\begin{equation}
z=1+\frac{2C_1}{L}=\frac{z_{\rm in}\sqrt{R}}{z_{\rm in}\sqrt{R}+(1-z_{\rm in})\sqrt{R_{\rm in}}}.
\end{equation}
Here, $n=6\nu_1/\nu_2$. So, we see that $n$ is inversely proportional to the viscous anisotropy $\nu_2/\nu_1$, which is related to $\alpha$ (Equation (\ref{alpha})). We use the boundary conditions, $l_x^{(0)}\rightarrow W_{\infty}$ as $R\rightarrow\infty$, and $l_x^{(0)}=W_{0,\rm in}$ at $R=R_{\rm in}$ for arriving at the above solution. The Equations (\ref{C1}) and (\ref{gl}) are used in arriving at the expressions for $z_{\rm in}$ and $z$.  

The solutions of the Equations (\ref{warp1}) and (\ref{warp2}) of first order in $a$ will take the following forms (Equation (33) of CB17) 
\begin{equation}\label{lx1}
l_x^{(1)}=\frac{W^r_{a,\rm in}\left(1-z^n\right)}{1-z_{\rm in}^n},
\end{equation}
and
\begin{eqnarray}\label{ly1}
l_y^{(1)} &=&\frac{W^i_{a,\rm in}\left(1-z^n\right)}{1-z_{\rm in}^n}+ \frac{8q}{\nu_2 (n^2-4)(1-z_{\rm in}^n)}.\left[-\frac{2z_{\rm in}\sqrt{R}
 +\sqrt{R_{\rm in}}(1-z_{\rm in})}{R(1-z_{\rm in})\sqrt{R_{\rm in}}}. \right. \nonumber
\\
 && \left. [(n-2)(W_{0,\rm in}-W_{\infty}z_{\rm in}^n)-(n+2)(W_{\infty}-W_{0,\rm in})z^n] \right. \nonumber
\\
&&\left. +\left(\frac{1-z^n}{1-z_{\rm in}^n}.\frac{1+z_{\rm in}}{R_{\rm in}(1-z_{\rm in})}\right). \right. \nonumber
\\
&&\left. [W_{0,\rm in}((n-2)+(n+2)z_{\rm in}^n)-2nW_{\infty}z_{\rm in}^n]~\right].
\end{eqnarray}
Here, $q=G^2M^2/c^3$. 
In order to obtain the above solutions, we use the following boundary conditions: 
$l_x^{(1)}(R_{\rm in})=W^r_{\rm a,in}$, $l_y^{(1)}(R_{\rm in})=W^i_{a,\rm in}$, 
and $l_x^{(1)}\rightarrow0$, $l_y^{(1)}\rightarrow0$ for $R\rightarrow\infty$ (which is reasonable as we do not expect the outer disk tilt to depend on the frame-dragging effect).

The solution (i.e., Equations (\ref{lx1}) and (\ref{ly1})) of the first order warped 
disk equations differ a little from the solution (Equation (33)) given in CB17. 
In particular, CB17 
focused on the disk inner edge tilt, and thus assumed the inner edge twist to be zero.
Here, we retain the disk inner edge twist term, and hence Equations (\ref{lx1}) and (\ref{ly1})
are more general than Equation (33) of CB17.

\subsubsection{Analytical solution of the second order warped disk equation}\label{next}
We now consider the Equation (\ref{warp}) of order $a^2$ for computing the expression of $l_x^{(2)}$. As mentioned before, these calculations were not performed in CB17. For this, we first make the Equation (\ref{warp1}) dimensionless using $R_g$ as the length scale, i.e. $R\rightarrow R/R_g$, and $C_1$ as the scale for angular momentum density. Upon using the above scheme, the dimensionless expression for $L$ can be given by
\begin{equation}
L\rightarrow L/C_1=C\sqrt{R}-2,
\end{equation}
where,
\begin{equation}
C=\frac{2z_{\rm in}}{z_{\rm in}-1}\frac{1}{\sqrt{R_{\rm in}}}.
\end{equation}
The Equation (\ref{warp1}) of order $a^2$ takes the following dimensionless form upon implementing the above scheme:
\begin{eqnarray}\label{nd}
\frac{\partial}{\partial R}\left[n l_x^{\left(2\right)}+R L \frac{\partial l_x^{\left(2\right)}}{\partial R}\right]&=&4 \xi \frac{L}{R^2}l_y^{\left(1\right)}\nonumber\\
&=&\frac{L}{R^2}\left[K.(1-z^n)+\frac{1}{\sqrt{R}}.\left(F-\frac{D}{\sqrt{R}}\right).\left(H-J z^n\right)\right]
\end{eqnarray}
where 
\begin{eqnarray}\label{constants}
H&=&(n-2)(W_{0,\rm in}-W_{\infty}z_{\rm in}^n),\ J=(n+2)(W_{\infty}-W_{0,\rm in}),\ \xi=\frac{c R_g}{\nu_2}, \nonumber\\
D&=&\frac{32\xi^2}{(n^2-4)(1-z_{\rm in}^n)},
\ F=D.\frac{2z_{\rm in}}{z_{\rm in}-1}.\frac{1}{\sqrt{R_{\rm in}}},
\ P=\frac{4\xi W_{a,\rm in}^i}{1-z_{\rm in}^n},\nonumber\\
K&=&P+D.\left[\frac{1+z_{\rm in}}{1-z_{\rm in}}.\frac{1}{R_{\rm in}.(1-z_{\rm in}^n)}.\lbrace W_{0,\rm in}((n-2)+(n+2)z_{\rm in}^n)-2nW_{\infty}z_{\rm in}^n\rbrace\right].
\end{eqnarray}
Integrating the Equation (\ref{nd}) we obtain
\begin{eqnarray}
&&n l_x^{(2)}+R L \frac{\partial l_x^{(2)}}{\partial R}=Q_1-\frac{D.H}{R^2}+\frac{2}{3}.\frac{
(C.D+2F).H}{R^{3/2}}- \frac{C.F.H-2K}{R}-\frac{2.C.K}{\sqrt{R}}\nonumber\\
&&-\frac{L^2z^n}{2R^2(n-4)(n-3)(n-2)}\times\left[(n-4)\lbrace F.J.(C\sqrt{R}-2n+4)-2.K.(n-3)\sqrt{R}\rbrace\sqrt{R}\right.\nonumber\\
&&\left.+D.J.\lbrace 12+2n^2-2n\left(5+C\sqrt{R}\right)+4C\sqrt{R}+C^2 R\rbrace\right],
\end{eqnarray}
where $Q_1$ is the integration constant. Now solving the above equation, we find
\begin{eqnarray}\label{lx2}
l_x^{(2)}&=&Q_2\left(\frac{\sqrt{R}}{L}\right)^n+\frac{z^n}{3R^2(n-2)(n-3)(n-4)}\times\left[-\sqrt{R}(n-4)\left(6K(n-3)(C\sqrt{R}-1)\sqrt{R}\right.\right.\nonumber\\
&+&\left.\left.F.J.\left(8-4n-3C\sqrt{R}+3nC\sqrt{R}-3C^2R\right)\right)+D.J.\left(-18+n^2\left(2C\sqrt{R}-3\right)+4C\sqrt{R}\right.\right.\nonumber \\
&+&\left.\left.3C^2R+3C^3R^{3/2}-3n\left(-5+2C\sqrt{R}+C^2 R\right)\right)\right]+\frac{1}{n(n+2)(n+3)(n+4)}.\frac{1}{6R^2}.\left[D.H\right.\nonumber\\
&.&\left.\left(n^3\left(-6+4C\sqrt{R}\right)+3C^4R^2+6n^2\left(-5+2C\sqrt{R}+RC^2\right)+n\left(-36+8C\sqrt{R}+6C^2R\right.\right.\right.\nonumber\\
&+&\left.\left.\left.6C^3R^{3/2}\right)\right)+(n+4)\sqrt{R}\left(6(n+3)\sqrt{R}\left(Q_1 R(n+2)-K\left(2n\left(-1+C\sqrt{R}\right)+C^2 R\right)\right)-F.\right.\right.\nonumber\\
&.&H .\left.\left.\left(n^2\left(-8+6C\sqrt{R}\right)+3C^3R^{3/2}+2n\left(-8+3C\sqrt{R}+3C^2R\right)\right)\right)\right],
\end{eqnarray}
where $Q_2$ is an integration constant. In order to obtain the expressions for $Q_1$ and $Q_2$, we assume the boundary condition, $l_x^{(2)}=0$ as $R\rightarrow\infty$, since we do not expect the outer edge of the disk to be affected by the frame-dragging effect. This condition, upon implementing on (\ref{lx2}), gives
\begin{equation}
Q_2=-\frac{(-C)^n.Q_1}{n}-\frac{(-C)^n}{6n(n+2)(n+3)(n+4)}.\left[3C^4.D.H-6C^2(n+3)(n+4)K-3.F.H.C^3(n+4)\right].
\end{equation}
We also consider $l_x^{(2)}$ to be zero at the inner edge $R=R_{\rm in}$. This is also reasonable when $a$ is small. Using this, we obtain the following expression of $Q_1$
\begin{eqnarray}
Q_1&=&\frac{n}{1-z_{\rm in}^n}\left[\frac{z_{\rm in}^n}{6n(n+2)(n+3)(n+4)}.\left(3C^4.D.H-6C^2(n+3)(n+4)K-3F.H.C^3(n+4)\right)\right.\nonumber\\
&-&\left.\frac{z^n}{3R_{\rm in}^2(n-2)(n-3)(n-4)}\times\left[-\sqrt{R_{\rm in}}(n-4)\left(6K(n-3)(C\sqrt{R_{\rm in}}-1)\sqrt{R_{\rm in}}\right.\right.\right.\nonumber\\
&+&\left.\left.\left.F.J.\left(8-4n-3C\sqrt{R_{\rm in}}+3nC\sqrt{R_{\rm in}}-3C^2R_{\rm in}\right)\right)+D.J.\left(-18+n^2\left(2C\sqrt{R_{\rm in}}-3\right)+4C\sqrt{R_{\rm in}}\right.\right.\right.\nonumber \\
&+&\left.\left.\left.3C^2R_{\rm in}+3C^3R^{3/2}_{\rm in}-3n\left(-5+2C\sqrt{R_{\rm in}}+C^2 R_{\rm in}\right)\right)\right]-\frac{1}{n(n+2)(n+3)(n+4)}.\frac{1}{6R_{\rm in}^2}.\left[D.H\right.\right.\nonumber\\
&.&\left.\left.\left(n^3\left(-6+4C\sqrt{R_{\rm in}}\right)+3C^4R^2_{\rm in}+6n^2\left(-5+2C\sqrt{R_{\rm in}}+R_{\rm in}C^2\right)+n\left(-36+8C\sqrt{R_{\rm in}}+6C^2R_{\rm in}\right.\right.\right.\right.\nonumber\\
&+&\left.\left.\left.\left.6C^3R^{3/2}_{\rm in}\right)\right)+(n+4)\sqrt{R_{\rm in}}\left(-6(n+3)\sqrt{R_{\rm in}}K\left(2n\left(-1+C\sqrt{R_{\rm in}}\right)+C^2 R_{\rm in}\right)\right.\right.\right.\nonumber\\
&-&F.H\left. .\left.\left.\left(n^2\left(-8+6C\sqrt{R_{\rm in}}\right)+3C^3R^{3/2}_{\rm in}+2n\left(-8+3C\sqrt{R_{\rm in}}+3C^2
R_{\rm in}\right)\right)\right)\right]\right].
\end{eqnarray}
Thus we derive the expressions for the quantities that are required to calculate the disk tilt angle up to linear order in $a$. One should note that all these expressions are valid up to a critical value of $a$ ($a_c$). Beyond $a_c$, the series expansion breaks down as the higher order terms become more dominant than the zeroth order term. We would discuss the regime of validity of the expression of the tilt angle (equation (\ref{beta_1})) in the section \ref{res2}.
\subsection{Numerical Solution of the warped disk equation}\label{num}
In this section, we discuss the setup for obtaining the numerical solution of the full warped disk Equations (\ref{warp1}) and (\ref{warp2}). We see that the Equations (\ref{warp1}) and (\ref{warp2}) form a set of second order coupled differential equations. In order to solve these equations, we would use the following boundary conditions:
\begin{eqnarray}\label{b1}
l_x(R_{\rm in})&=&\beta_i \cos(\gamma_i),\ l_y(R_{\rm in})=\beta_i \sin(\gamma_i),
\end{eqnarray}
and 
\begin{eqnarray}\label{b2}
l_x(R_{f})&=&\beta_f, \ l_y(R_{f})=0,
\end{eqnarray}
where $R_f$, $\gamma_i$, $\beta_i$ and $\beta_f$ are the outer edge radius, twist angle at the inner boundary, the tilt angle at the inner edge, and the outer edge of the disk, respectively. We assume the twist angle to be zero at the outer edge of the disk as the effect of LT precession on the outer boundary is negligible. Hence, $l_y$ can be assumed to be zero at the outer edge (see the expression of twist angle below Equation (\ref{warp})). As mentioned before, $R_{\rm in}$ is the ISCO radius ($R_{\rm ISCO}$) of a prograde disk, which for a Kerr black hole takes the following form (\cite{Teukolsky})
\begin{equation}
R_{\rm in}=R_{\rm ISCO}=\left[3+Z_2-\lbrace{\left(3-Z_1\right)\left(3+Z_1+2Z_2\right)\rbrace}^{1/2}\right],
\end{equation}
where $Z_1=1+(1-a^2)^{1/3} \left[(1+a)^{1/3}+(1-a)^{1/3}\right]$, and $Z_2=\left(3 a^2+Z_1^2\right)^{1/2}$ ($R_{\rm in}$ has been made dimensionless using $R_g$). So, in our construction the solution of the coupled differential equations depend upon the values of $\beta_i$, $\beta_f$, and $\gamma_i$ through the boundary conditions. In this paper, we fix the values of $\beta_f$ and $\gamma_i$, and keep the value of $\beta_i$ as a free parameter. The above inner boundary condition (\ref{b1}) can lead to nonzero viscous torques at the inner edge, when the surface density is nonzero at the inner boundary. We discuss such a possibility in the next section. 

We derive the dimensionless versions of the Equations (\ref{warp1}) and (\ref{warp2}) using the scheme mentioned in the previous subsection, and obtain 
\begin{eqnarray}\label{noeq}
R\frac{\partial^2 l_x}{\partial R^2}+\left[(n+1)\frac{C_1}{L}+3/2\right]\frac{\partial l_x}{\partial R}&=&4 a \xi \frac{l_y}{R^2},\\ \nonumber
R\frac{\partial^2 l_y}{\partial R^2}+\left[(n+1)\frac{C_1}{L}+3/2\right]\frac{\partial l_y}{\partial R}&=&-4 a \xi \frac{l_x}{R^2}.
\end{eqnarray}
We now solve the above Equations (\ref{noeq}) subjected to the boundary conditions mentioned in (\ref{b1}) and (\ref{b2}). We see from the above equations that the ratio $aM/\nu_2$ ($\nu_2$ and $M$ appear in $\xi$) captures the essence of the problem.

\subsection{Viscous torques at the inner boundary}\label{vis}

In this section, we discuss the implications of our inner boundary conditions 
on viscous torques.
As discussed in section \ref{formalism}, there are two viscous torques acting on 
the accretion disk, one acting perpendicular to the disk ($\textbf{G}_1$), 
and the other in the plane of the disk ($\textbf{G}_2$). The first torque can be given by \citep{Pap83,Nixon}
\begin{equation}\label{per}
\textbf{G}_1=2\pi R\nu_1 \Sigma R \Omega^{'} R \textbf{l}.
\end{equation}
Using the expressions of $\Omega$ (Equation (\ref{omega})) and angular momentum density (Equation (\ref{L})) for a Keplerian disk, we can arrive at the following expression of $\textbf{G}_1$, 
\begin{eqnarray}
\textbf{G}_1&=&-3\pi \nu_1 \Sigma \sqrt{GM} R^{1/2} \textbf{l}\nonumber\\
&=&-3\pi \nu_1 L(R)  \textbf{l}.
\end{eqnarray}
Since in our formalism $L(R)$ and $\Sigma(R)$ are nonzero at the inner edge (see the Equations (\ref{b1}), (\ref{sig}) and (\ref{gl}), and the text after Equation (\ref{C2})), we see from the above equation that the $x$ and $y$ components of the torque $\textbf{G}_1$ vanish (although, the $z$ component of $\textbf{G}_1$ remains nonzero) at the inner boundary only when the tilt angle at the disk inner edge is zero.
In this paper, we also consider the scenario, in which the inner edge tilt angle is nonzero. In such a case, even the $x$ and $y$ components of the torque $\textbf{G}_1$ do not vanish at the inner boundary.

Now we discuss the implication of the disk inner edge tilt on the torque $\textbf{G}_2$, which arises only 
if the disk is warped. 
This torque can be given by \citep{Pap83,Nixon}
\begin{equation}\label{par}
\textbf{G}_2=\pi R\nu_2\Sigma \Omega R^2 \frac{\partial \textbf{l}}{\partial R}.
\end{equation}
Again, using the Equations (\ref{omega}) and (\ref{L}) for $\Omega$ and angular momentum density 
respectively for a Keplerian disk, we can arrive at the following expression of $\textbf{G}_2$, 
\begin{eqnarray}
\textbf{G}_2&=&\pi R\nu_2\Sigma(R) \sqrt{GM} R^{1/2}\frac{\partial \textbf{l}}{\partial R}\nonumber\\
&=&\pi \nu_2 R L(R) \frac{\partial \textbf{l}}{\partial R}.
\end{eqnarray}
Since $\Sigma(R)$ and $L(R)$ do not vanish at $R_{\rm in}$ in our formalism (see above),
the $x$ and $y$ components of $\textbf{G}_2$ are zero at the inner edge only when the corresponding components of $\partial \textbf{l}/\partial R$ vanish at the inner boundary.
Note that $\partial \textbf{l}/\partial R$ at $R_{\rm in}$ is zero when the inner part of the disk is aligned
with the black hole spin equator, but is nonzero for a non-aligned inner disk, even when the inner edge tilt angle is zero.

We stress that the inner boundary conditions we choose both for analytical and numerical calculations are 
quite general, and we do not invoke any additional constraints on the inner edge, like that considered 
in \cite{Pringle92} and \cite{Lodato} to maintain an accreting torque-free inner boundary. The nonzero viscous torque at the inner boundary has been discussed earlier in \cite{Kulkarni},\cite{Penna}, \cite {Zhu}, and \cite{McClintock}.

\section{Results and Discussion}\label{result}
In this section, we first discuss the quantitative differences between our analytical radial profiles of the disk tilt angle, and the same mentioned in CB17. Then, we explore the validity of the numerical results by comparing the same with that obtained from analytical calculations, and up to what extent the analytical results remain relevant. Finally, we present numerically computed radial profiles of the disk tilt angle for various sets of parameter values, and discuss the implications of the results. 
\begin{figure}
\begin{center}
\includegraphics[width=0.49\textwidth]{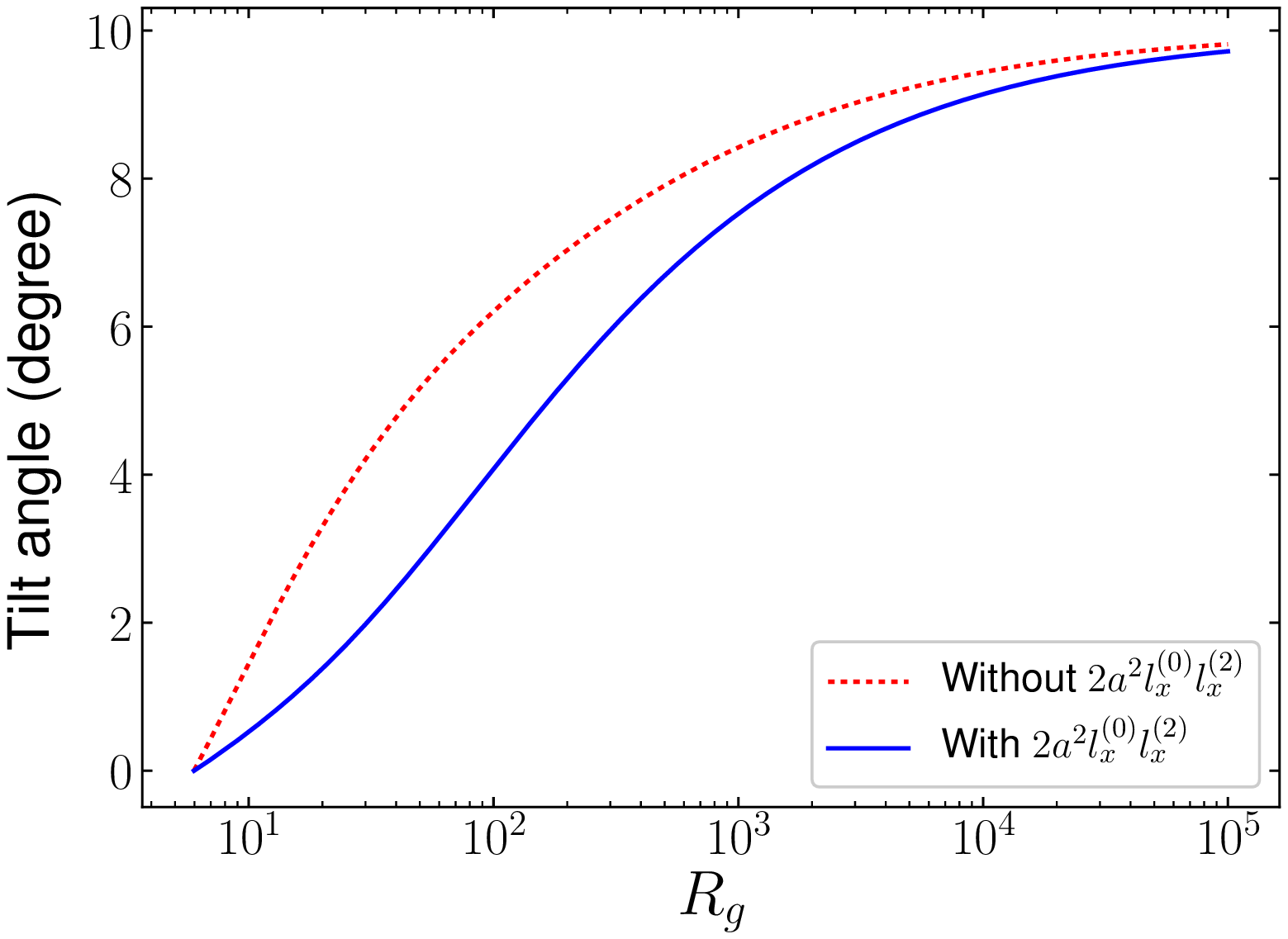}
\includegraphics[width=0.49\textwidth]{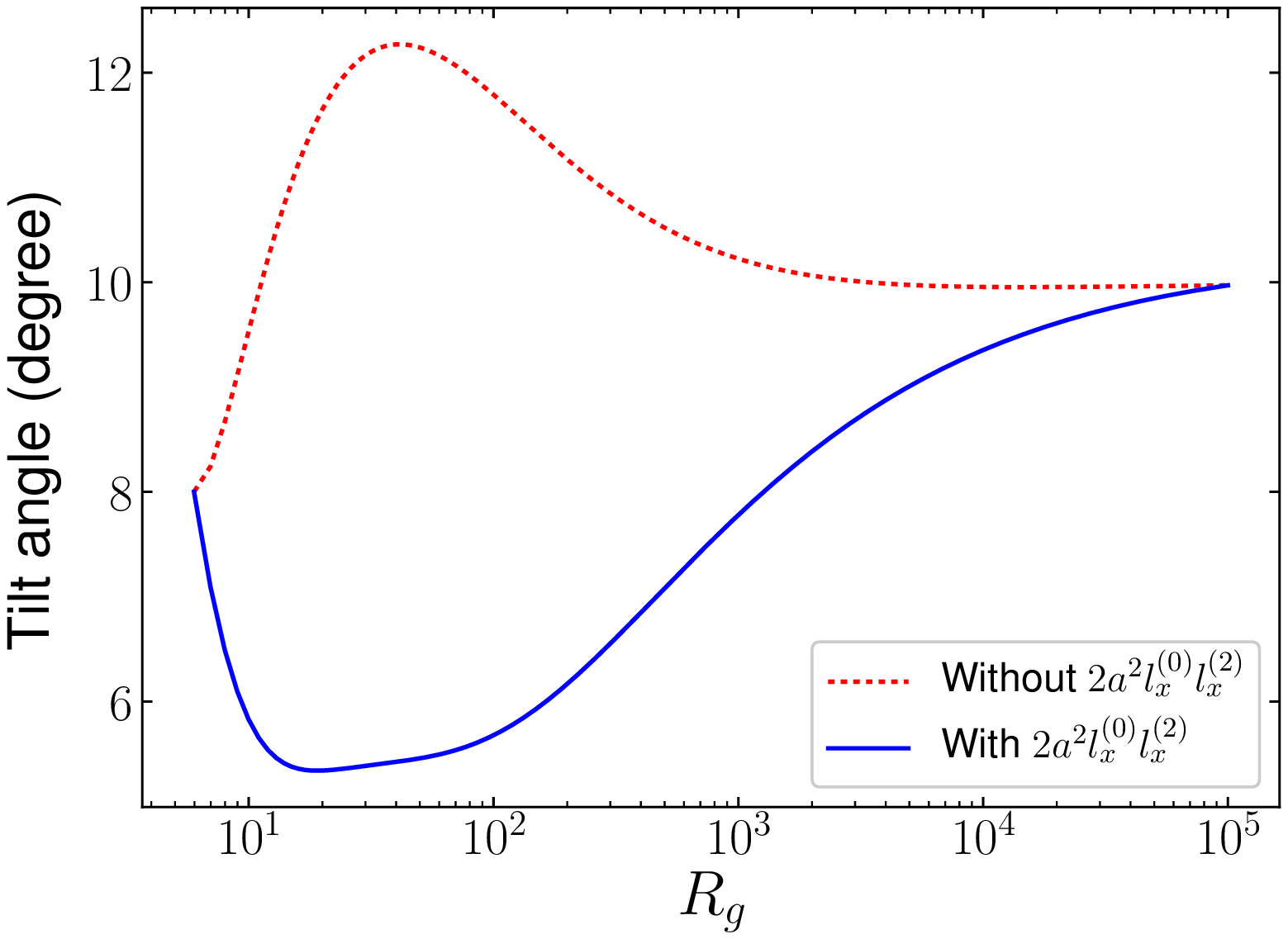}
\caption{Radial profiles of the disk tilt angle, including and excluding the term  $2a^2l_x^{(0)}l_x^{(2)}$ as given in the 
Equation (\ref{beta_1}). The parameter values ($a=0.007$, $M=10M_{\odot}$, $n=6$, $\nu_1=\nu_2=10^{14}$ $\rm cm^2$ $\rm s^{-1}$ and $z_{\rm in}=0.75$) are chosen in accordance with CB17. $W_{0,\rm in}=0^{\circ}$ for the left panel and $=8^{\circ}$ for the right panel (see section \ref{res1} for details).}
\end{center}
\label{fig1}
\end{figure}

In order to discuss the behavior of analytically as well as numerically obtained radial profiles of disk tilt angle, we have to choose suitable values of different parameters relevant to the astronomical scenario we are considering. Since in this paper we are mostly interested in the case of Galactic accreting black holes, we choose the mass of the black hole in the range of $5-15 M_{\odot}$ (\cite{Fragos}). 
We consider the viscosity $\nu_2$ in the range $10^{14}-10^{15}$ $\rm cm^2$ $\rm s^{-1}$ (\cite{Frank}), a range of $0.0012-0.94$ for $n$, which translates to the range $0.01-0.4$ for $\alpha$ (\cite{King}) (Equation (\ref{alpha}) has been used for calculating the range of $\alpha$), and a range of $0.3-0.75$ for $z_{\rm in}$ (as in our formalism $\Sigma_{\rm in}>\Sigma_{\infty}$). We set the inner edge twist and outer edge tilt to $5^\circ$ and $10^\circ$, respectively, throughout the paper. We consider $\beta_i$ as a free parameter, and use the range $0^\circ-10^\circ$ for the purpose of demonstration. Also in this paper we take into account only the case of prograde rotation $(a>0)$. 
\begin{figure}[h]
\begin{center}
\includegraphics[width=0.49\textwidth]{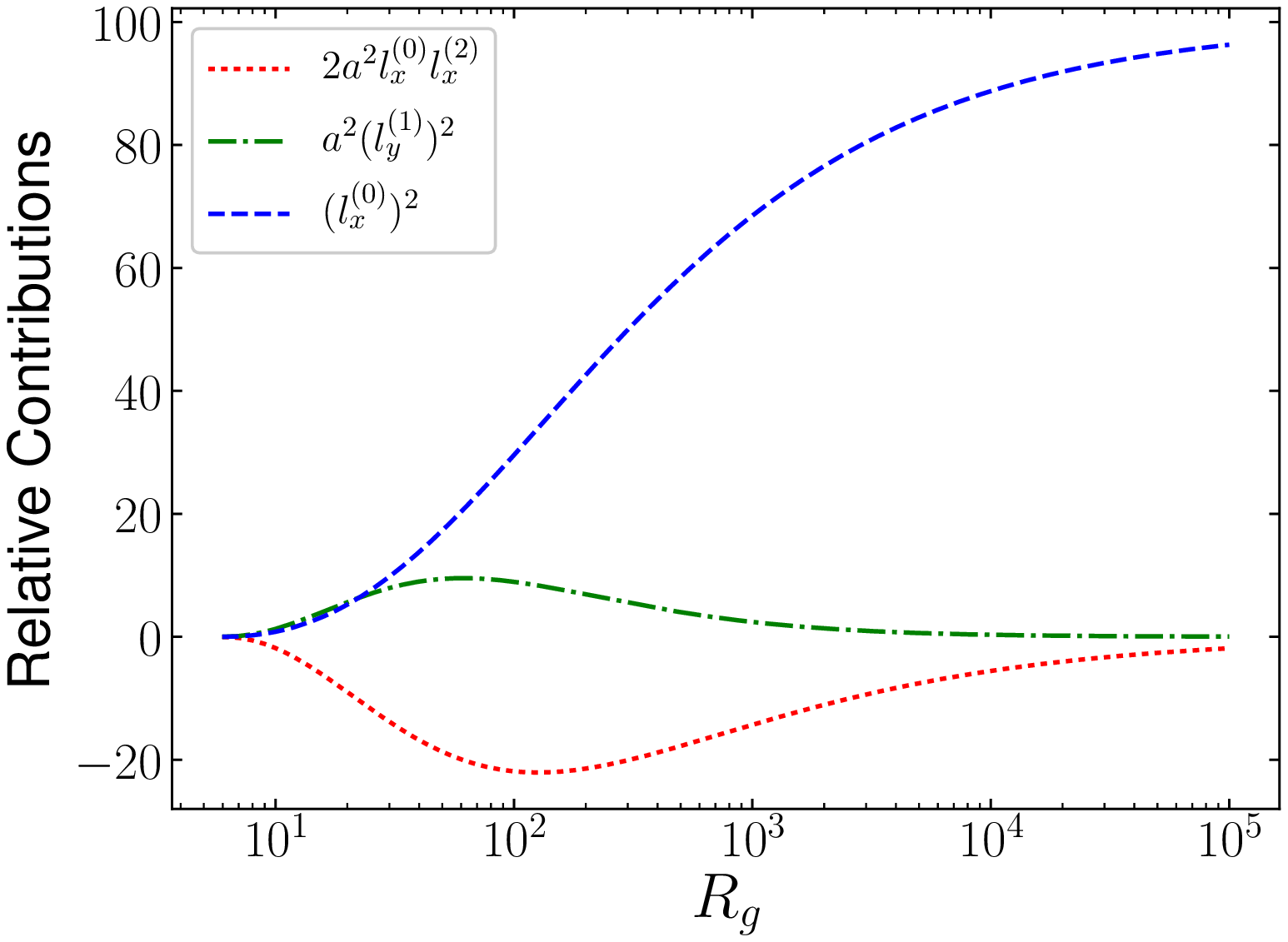}
\includegraphics[width=0.49\textwidth]{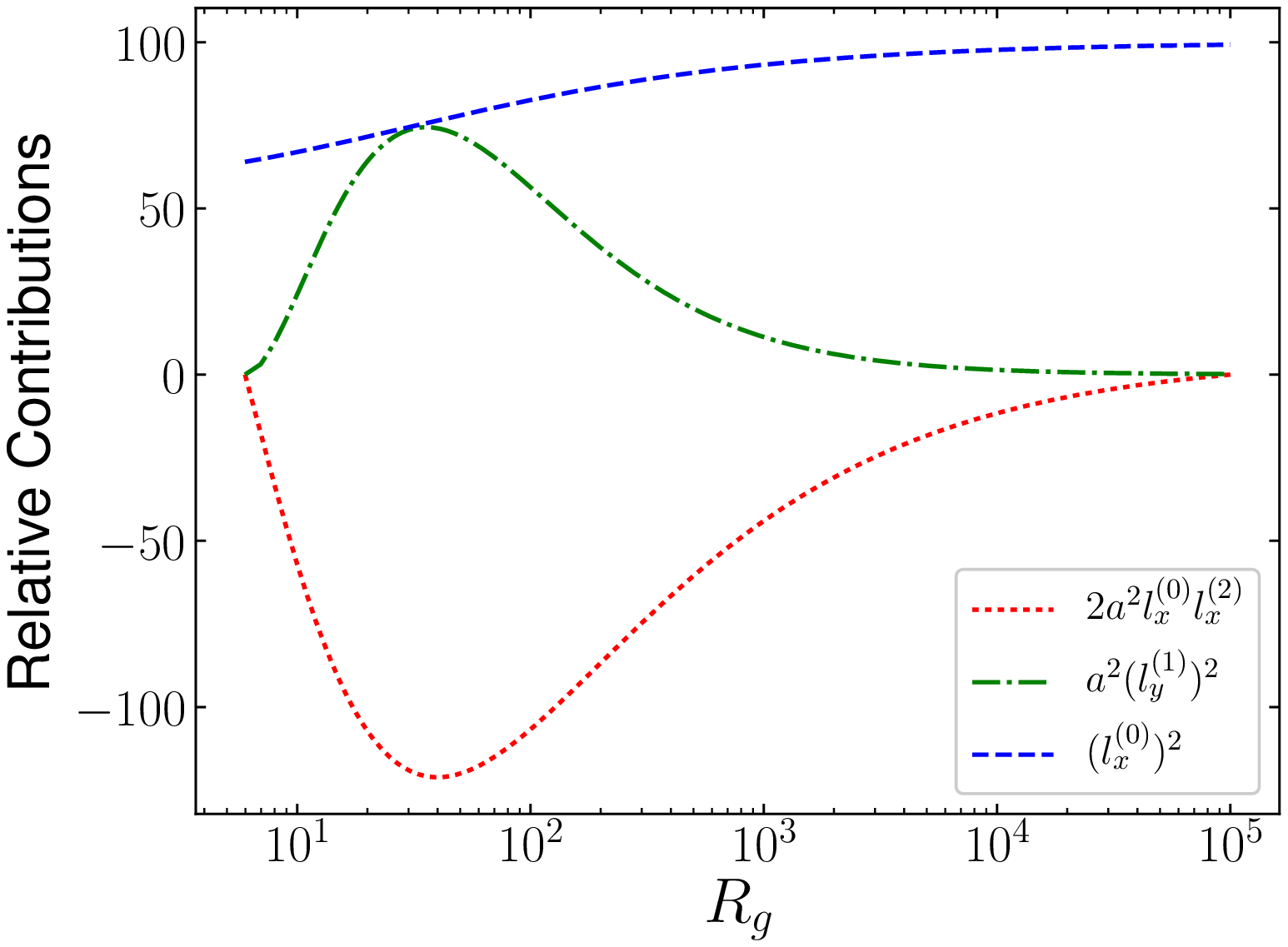}
\caption{Relative contributions of the terms present in the expression of the tilt angle (Equation \ref{beta_1}) for two different initial tilt angles (also considered in CB17 for demonstration). We use the same parameter values that have been used in Figure (\ref{fig1}). All the quantities (e.g. $l_x^{(2)}$) in each term are expressed in degrees. The disk inner edge tilt angle in the left panel and right panel are $0^{\circ}$ and $8^{\circ}$ respectively (see section \ref{res1} for details).}
\label{fig2}
\end{center}
\end{figure}
\subsection{Comparison between analytical results obtained in CB17 and our analysis}\label{res1}
As mentioned earlier in the section \ref{an}, the expression of the tilt angle obtained in CB17 is incomplete, as the term $2a^2l_x^{(0)}l_x^{(2)}$ was ignored. In order to appreciate the significance of this term, we compare the radial profiles of the disk tilt angle in Figure \ref{fig1} obtained by including this term, and without including this term. We also consider same range of the parameters used in CB17 for this purpose, and perform the comparison for two different inner edge tilt angles ($W_{0,\rm in}$). The mismatch between the results can be seen to be severe for nonzero $W_{0,\rm in}$, whereas for the zero inner edge tilt angle, the tilt angle profile remains qualitatively the same to some extent. We find that the inclusion of the term $2a^2l_x^{(0)}l_x^{(2)}$ replaces the hump part of the radial profile as reported in CB17  with a dip. To probe this difference closely, we have compared the contributions of the term $2a^2l_x^{(0)}l_x^{(2)}$ along with the terms $a^2(l_y^{(1)})^2$ and $(l_x^{(0)})^2$ in Figure \ref{fig2}, for the same parameter values used in Figure \ref{fig1}, to check their relative dominance. We see that the term $2a^2l_x^{(0)}l_x^{(2)}$ completely suppresses the effect of the term involving $l_y^{(1)}$, which is responsible for the hump. This is exactly why we find a dip instead of a hump reported in CB17. The results match qualitatively for the case, when the tilt angle at the inner edge is zero, as the higher order terms are quite smaller than the zeroth order term for this particular scenario than the cases for which inner edge tilt angle is nonzero. We also check that the other two terms $a^2\left(l_x^{\left(1\right)}\right)^2$ and $2al_x^{(1)}l_x^{(0)}$ (see the terms present in equation (\ref{beta_1})), which have been set to zero for the demonstration purposes in CB17, do not change the result as their contributions are quite small compared to the ones mentioned in Figure \ref{fig2}.
\begin{figure}[h!]
\includegraphics[width=0.49\textwidth]{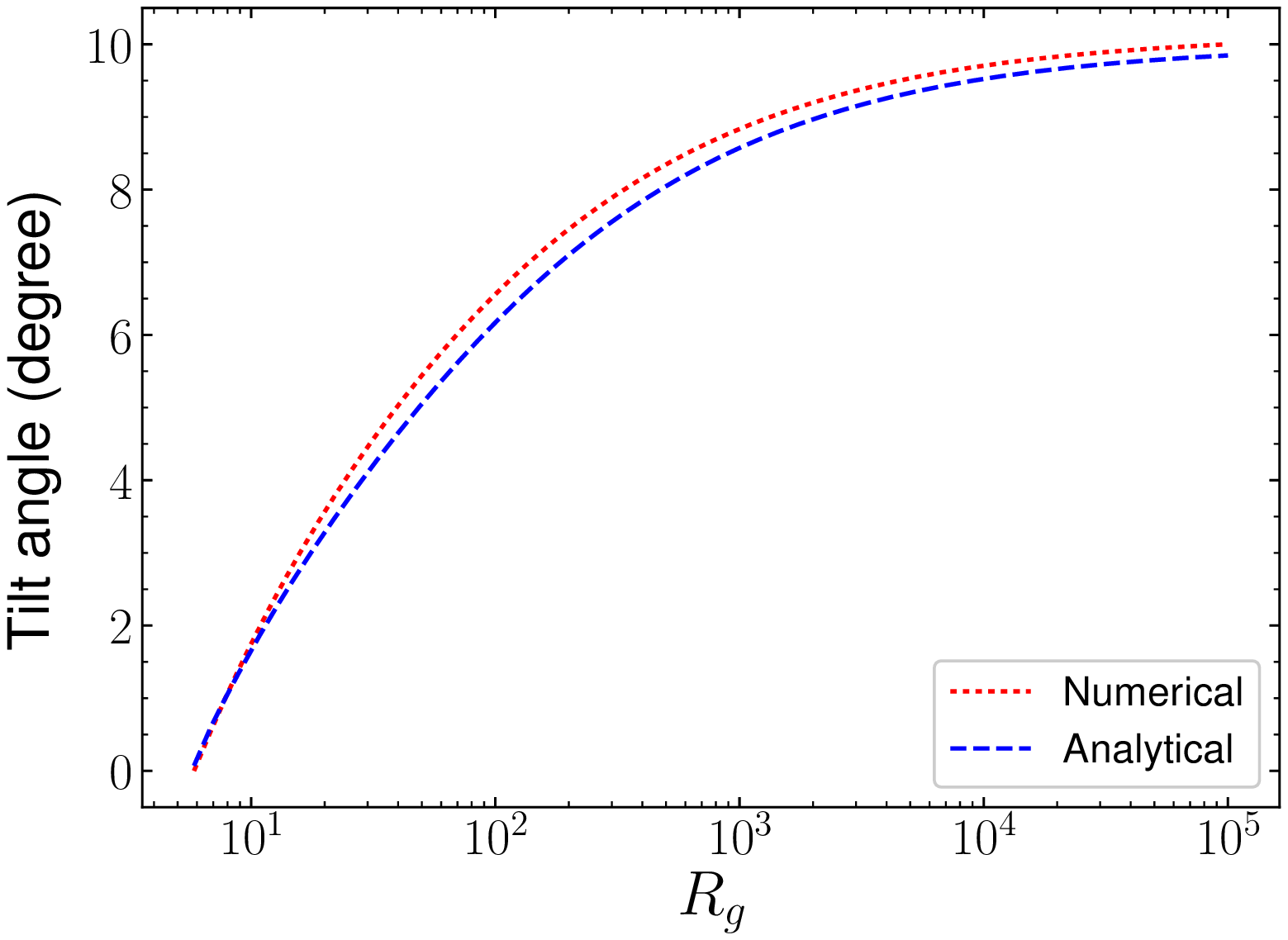}
\includegraphics[width=0.49\textwidth]{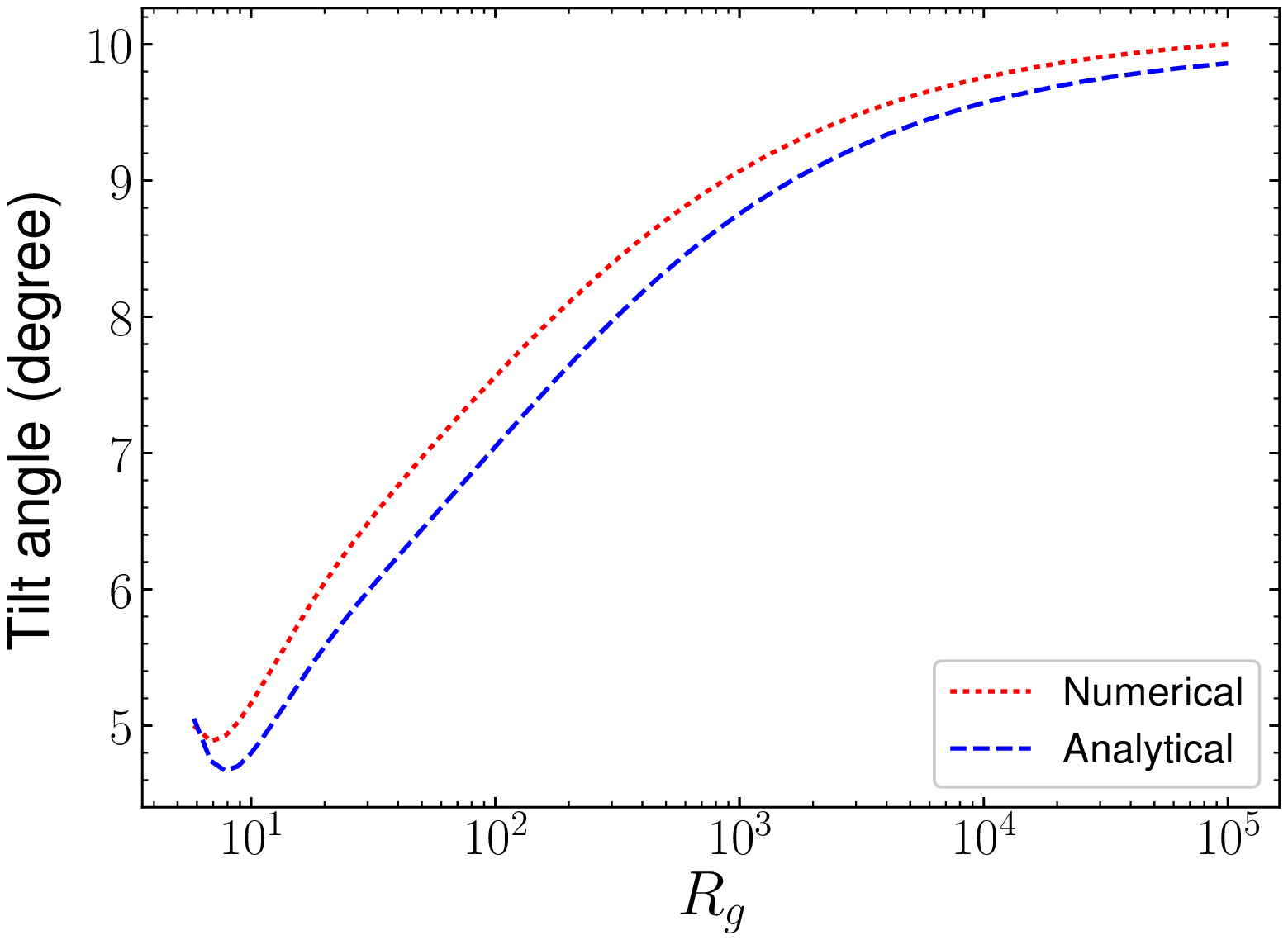}
\caption{Radial profiles of the disk tilt angle obtained using the analytical equation (\ref{beta_1}), and by numerically solving the warped disk equations (\ref{noeq}) for $a=0.05$ and two different disk inner edge tilt angles ($0^\circ$ for the left panel and $5^\circ$ for the right panel). The other parameter values are $M=10M_{\odot}$, $W_{a,\rm in}^r=1^{\circ}$, $W_{a,\rm in}^i=1^{\circ}$, $\nu_2=10^{15}$ $\rm cm^2 $ $\rm s^{-1}$, $n=0.25$ and $z_{\rm in}=0.75$. In the case of the numerical result, we have to specify the disk inner edge twist angle, and we choose $\gamma_i=1^{\circ}$ (see section \ref{res2} for details).}
\label{fig3}
\end{figure}
\begin{figure}[h!]
\includegraphics[width=0.49\textwidth]{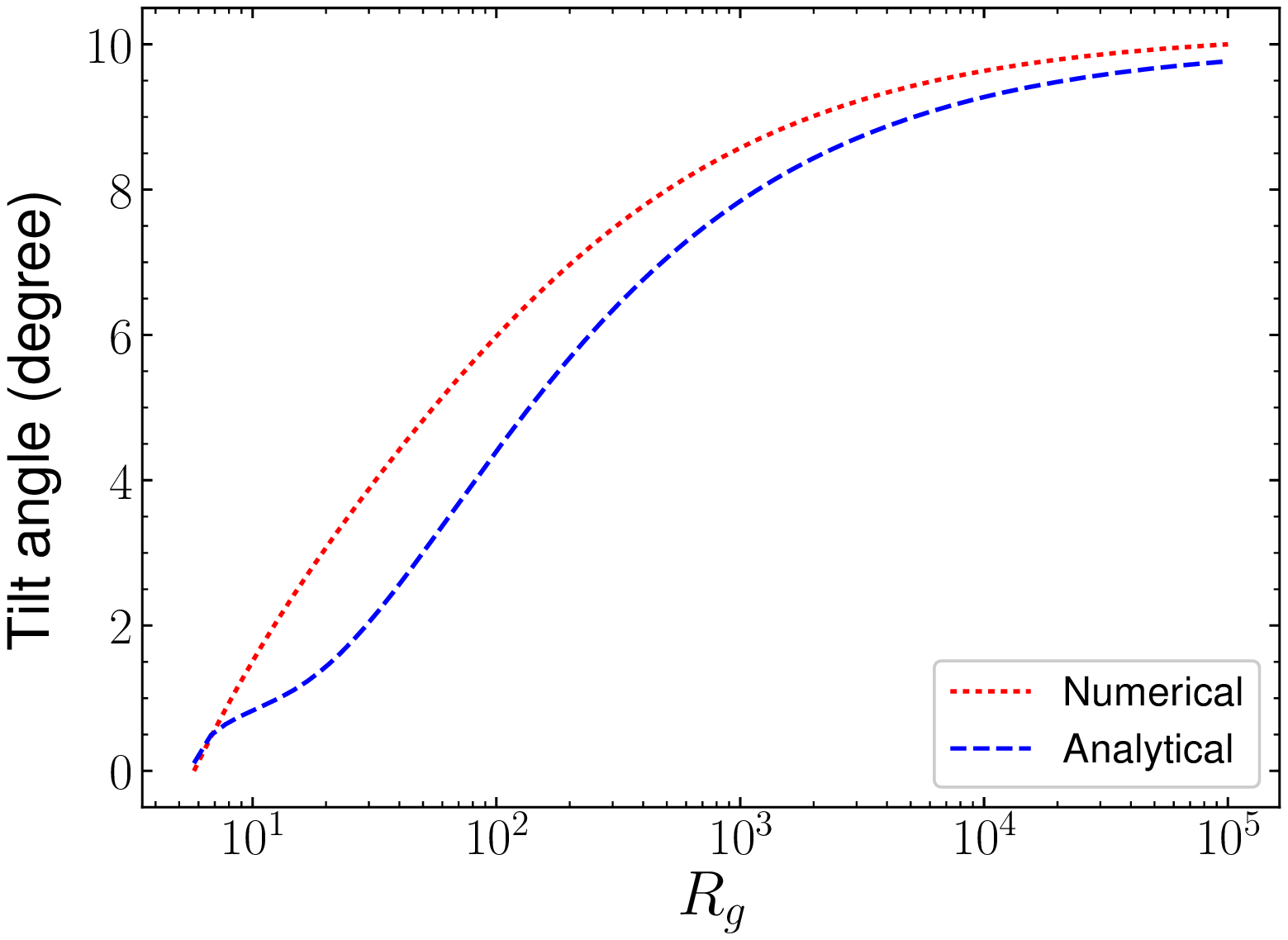}
\includegraphics[width=0.49\textwidth]{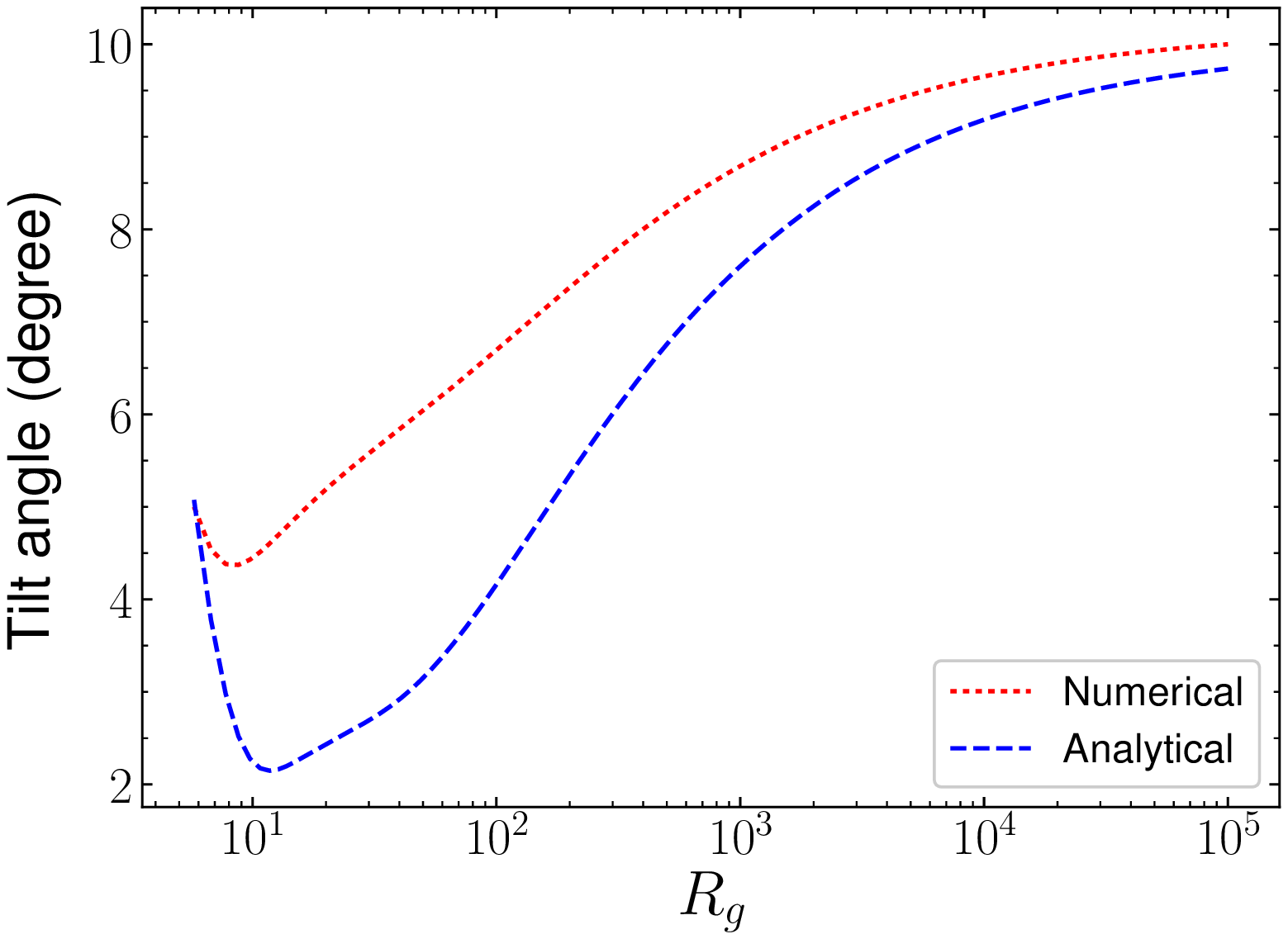}
\caption{Same as Figure \ref{fig3}, with the exception of $a=0.08$ (see section \ref{res2} for the discussion related to the panels).}
\label{fig4}
\end{figure}
\begin{figure}[h!]
\begin{center}
\includegraphics[width=0.49\textwidth]{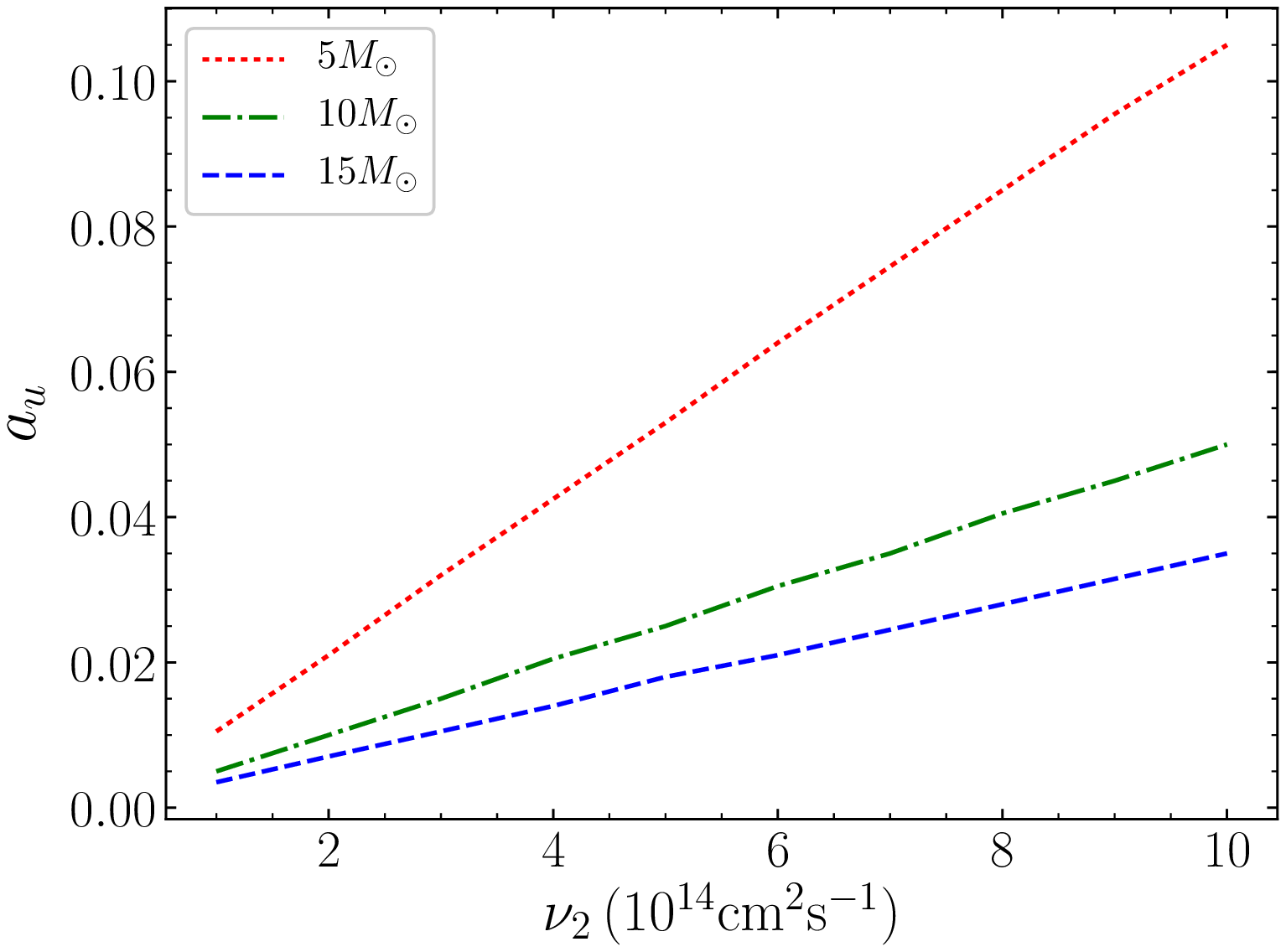}
\caption{The upper limit of the Kerr parameter $a_u$ (up to which the analytical result matches with the numerical result within $10\%$) for three different masses of the black hole against $\nu_2$. Other parameter values are $\beta_i=5^{\circ}$, $n=0.25$, and $z_{\rm in}=0.75$. The curves depend very weakly upon these latter mentioned parameters (see section \ref{res2} for details).}
\label{fig5a}
\end{center}
\end{figure}
\subsection{Comparison between analytical and numerical results}\label{res2}
Now we compare the radial profile of the disk tilt angle obtained using the analytical expression (\ref{beta_1}) with the same obtained by numerically solving the warped disk Equations (\ref{noeq}). We also like to find the critical value of the Kerr parameter $a_c$, up to which the analytical expression is valid, and the upper limit of the Kerr parameter $a_u$, up to which the analytical result matches with the numerical result within $10\%$. The critical value of the Kerr parameter $a_c$ depends strongly on the values of $\nu_2$ and $M$ as the dominance of the terms $l_y^{(1)}$ and $2a^2l_x^{(0)}l_x^{(2)}$ (main contributor to $\beta^{(1)}$ apart from $l_x^{(0)}$, see the section \ref{res1} for details) are controlled by $\nu_2$ or $M$ through $\xi$ or $q$ (see the Equations (\ref{ly1}) and (\ref{lx2})). When $\nu_2=10^{15}$ $\rm cm^2$ $\rm s^{-1}$, the analytical expression remains valid up to $a\sim0.08$, whereas for $\nu_2=10^{14}$ $\rm cm^2$ $\rm s^{-1}$ it remains defined up to $a\sim0.008$ for $M=10M_{\odot}$. Beyond this value of $a_c$, the higher order terms become more dominant than the zeroth order term. The upper limit of the Kerr parameter $a_u$ also similarly depends upon $\nu_2$ and $M$. When $\nu_2=10^{15}$ $\rm cm^2$ $\rm s^{-1}$, the numerically obtained radial profile matches well (within $\sim 10\%$) with the same obtained using analytical expression (\ref{beta_1}) up to $a\sim0.05$ (see Figure \ref{fig3}), beyond which the mismatch becomes prominent (see Figure \ref{fig4} where $a=0.08$). 
\begin{figure}[h!]
\includegraphics[width=0.49\textwidth]{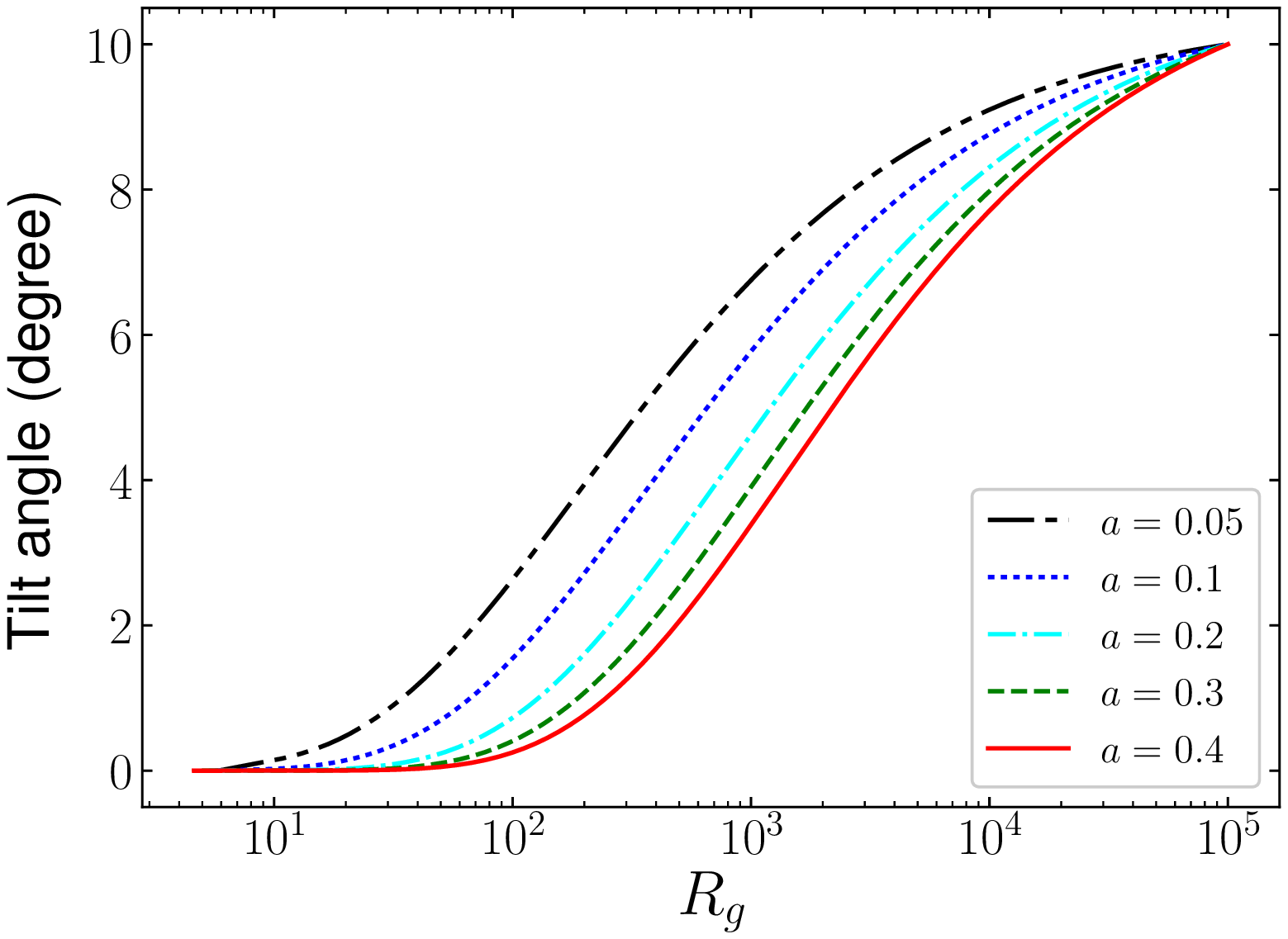}
\includegraphics[width=0.49\textwidth]{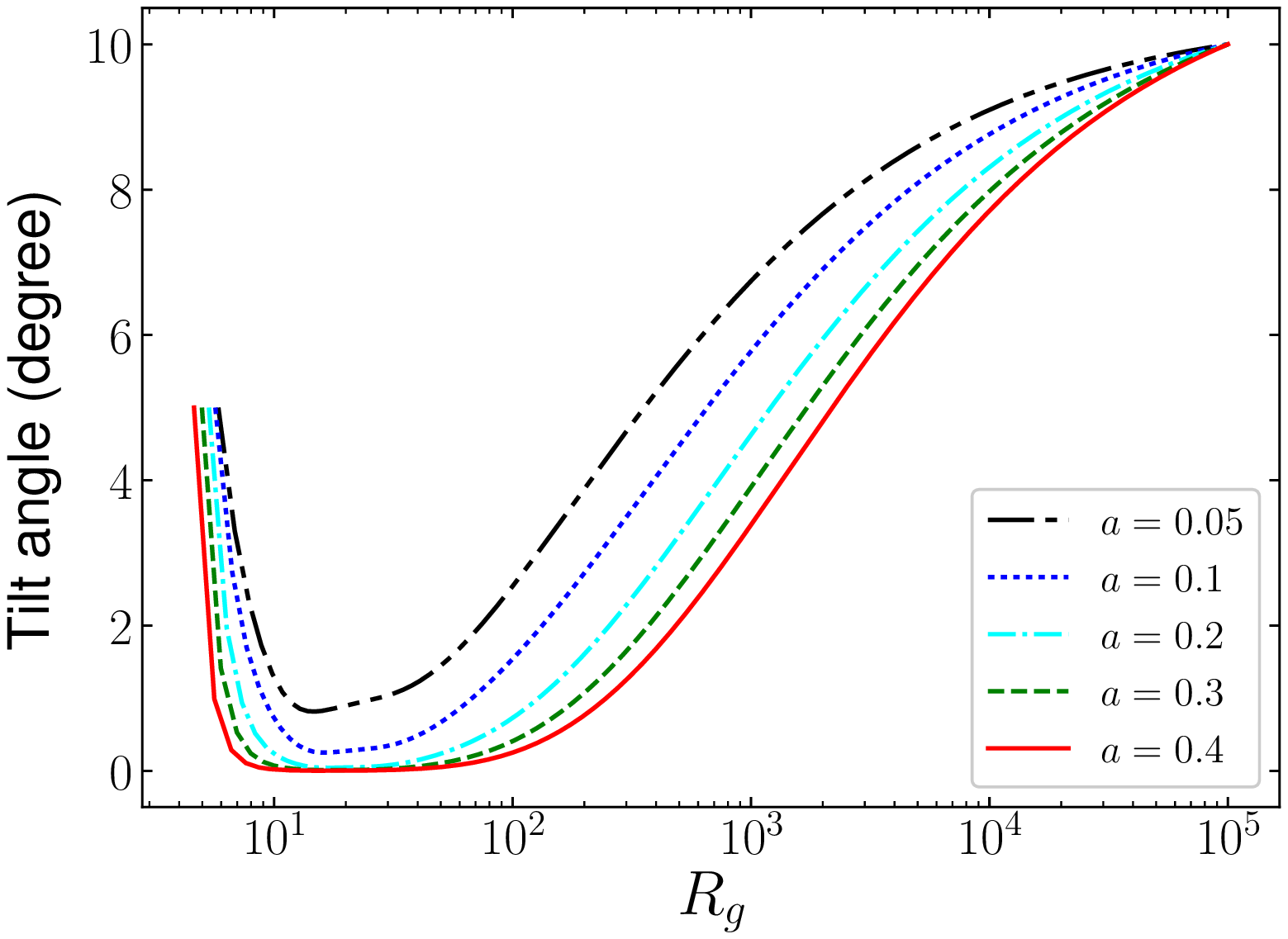}
\caption{Radial profiles of the disk tilt angle for different values of the Kerr parameter $a$ ($\beta_i=$ $0^{\circ}$ for the left panel and $5^{\circ}$ for the right panel) for $\nu_2=10^{14}$ $\rm cm^2$  $s^{-1}$. The other parameter values are $M=10M_{\odot}$, $n=0.25$,  and $z_{\rm in}=0.75$ (see section \ref{res3} for details).}
\label{fig5}
\end{figure}
For $\nu_2=10^{14}$ $\rm cm^2$  $s^{-1}$ and $M=10M_{\odot}$, the analytical result matches with the numerical result up to $10\%$ for $a\lesssim0.005$. To probe this limit more closely, we plot $a_u$ against $\nu_2$ for three different values of $M$ (Figure \ref{fig5a}), and this plot elucidates the relation among these three quantities.  

We see that the analytical result is valid only for very small $a$ values. Therefore, numerical computation, which is valid also for large $a$ values, 
is required to confront the observations. The analytical calculation, however, is important to gain an insight into the system, and to test the numerical results for the same parameter values.
\begin{figure}[h]
\includegraphics[width=0.49\textwidth]{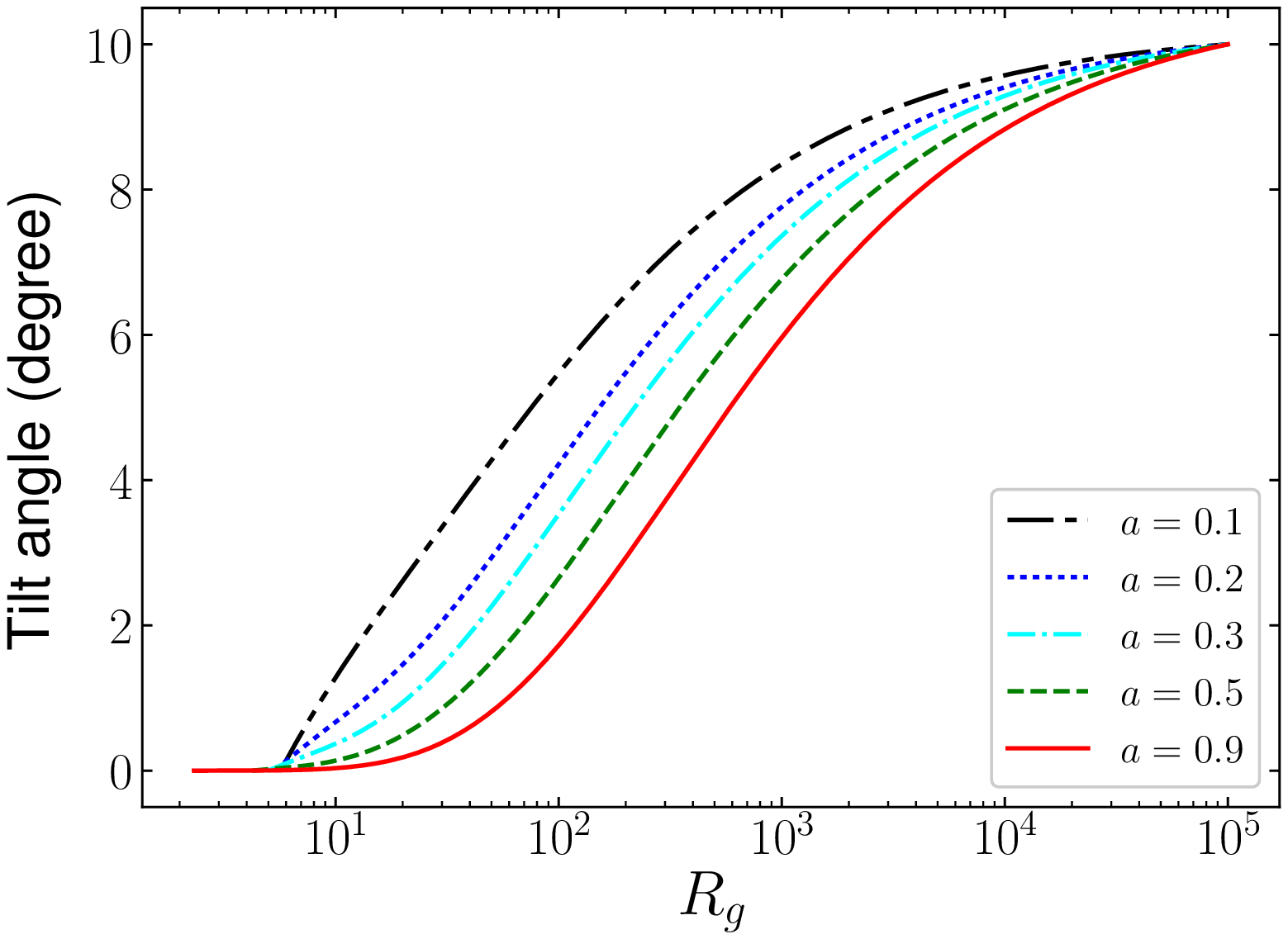}
\includegraphics[width=0.49\textwidth]{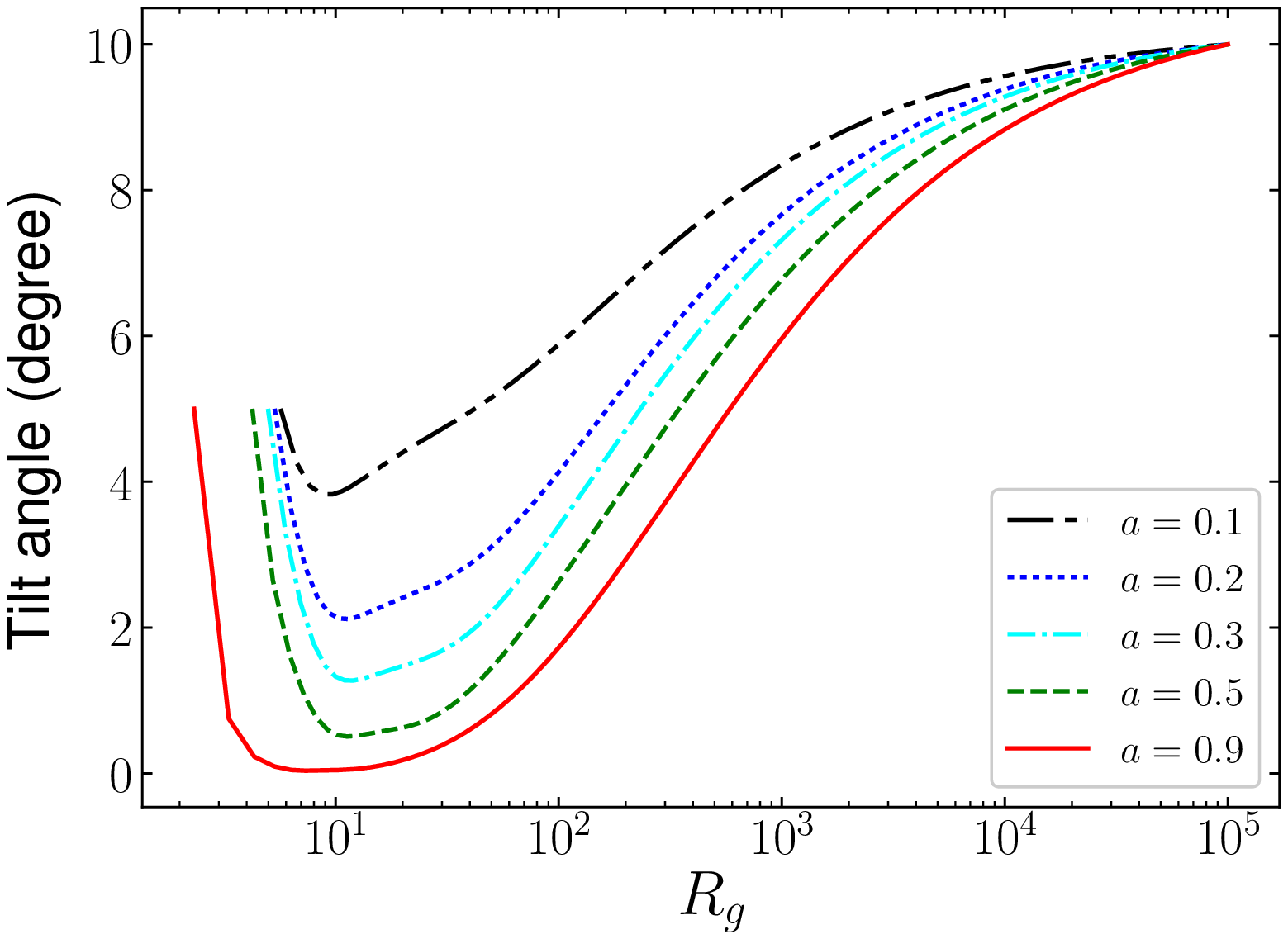}
\caption{Same as Figure \ref{fig5}, with the exception of $\nu_2=10^{15}$ $\rm cm^2$ s$^{-1}$ (see section \ref{res3} for details).}
\label{fig6}
\end{figure}
\begin{figure}[h]
\begin{center}
\includegraphics[width=0.49\textwidth]{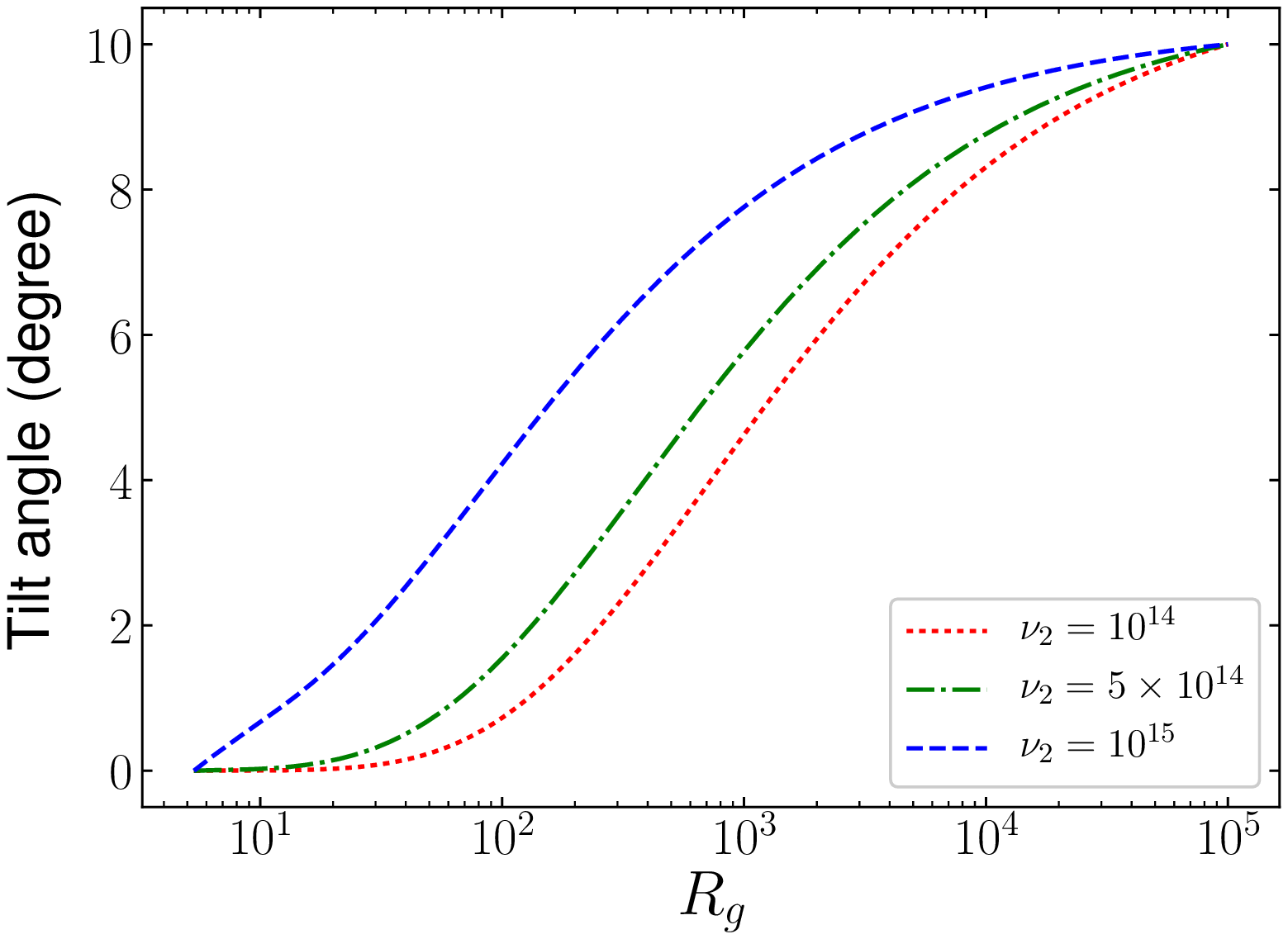}
\includegraphics[width=0.49\textwidth]{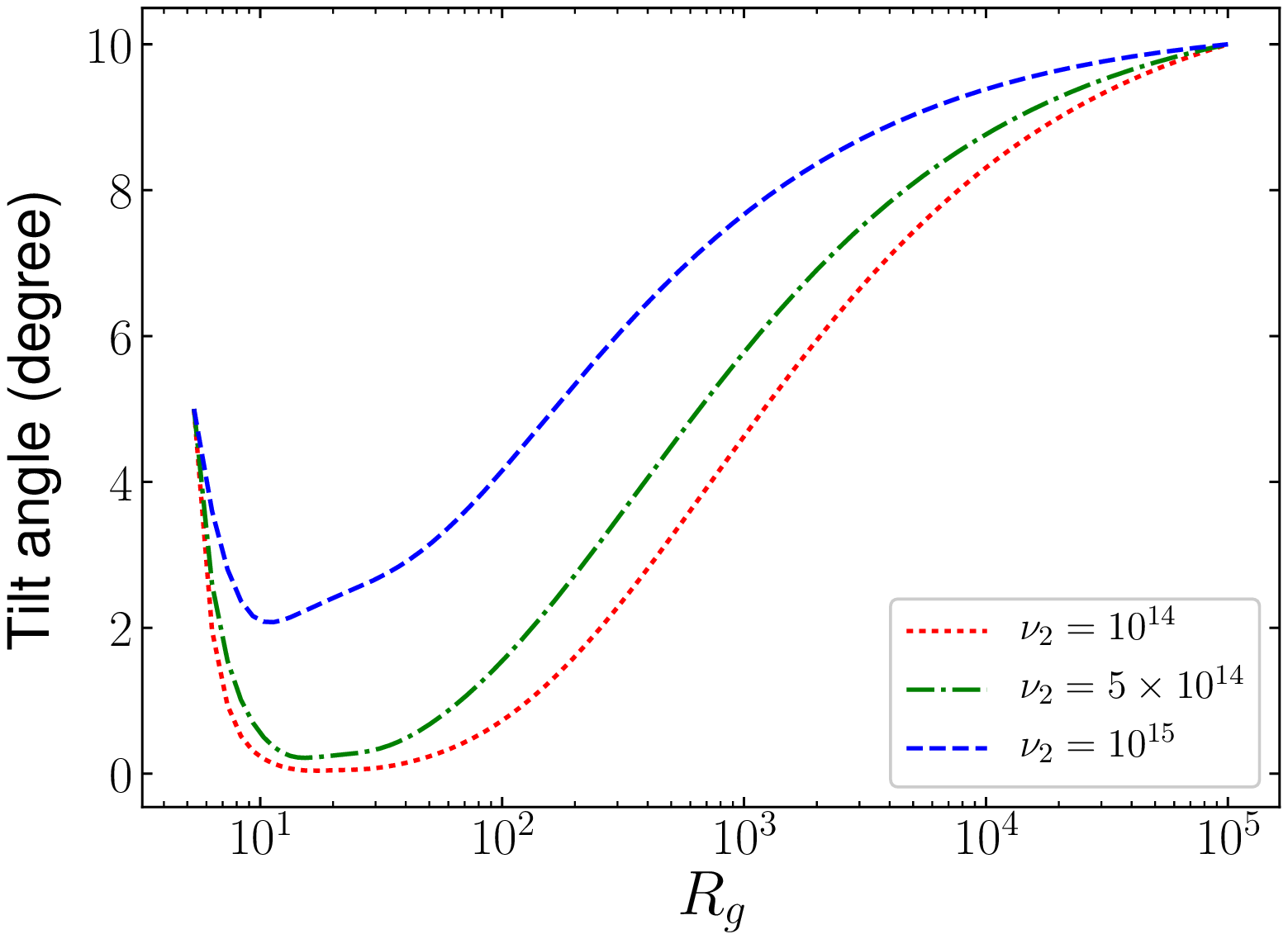}
\caption{Radial profiles of the disk tilt angle for different values of $\nu_2$ ($\beta_i=$ $0^{\circ}$ for the left panel and $5^{\circ}$ for the right panel). The other parameter values are $a=0.2$, $M=10M_{\odot}$, $n=0.25$, and $z_{\rm in}=0.75$ (see section \ref{res3} for the details).}
\label{fig7a}
\end{center}
\end{figure}

\subsection{Numerically computed radial profiles of the disk tilt angle and their implications}\label{res3}
We investigate in detail the behavior of the tilt angle radial profile as a function of the parameters $a$, $\beta_i$, $\nu_2$ and $n$. As discussed earlier, the interplay between the LT torque (controlled by the parameters $M$, $a$ and $\beta_i$) and viscous torque ($\bf{G_2}$) in the plane of the disk (controlled by $\nu_2$) strongly decides the tilt profile of the disk. 
A higher Kerr parameter ($a$) value implies a stronger LT torque, for fixed values of other parameters. 
Hence, for a higher $a$, the angle between the inner disk angular momentum vector 
and the black hole spin axis is smaller (see Figure \ref{fig5} and Figure \ref{fig6}).  
\begin{figure}[h]
\begin{center}
\includegraphics[width=0.49\textwidth]{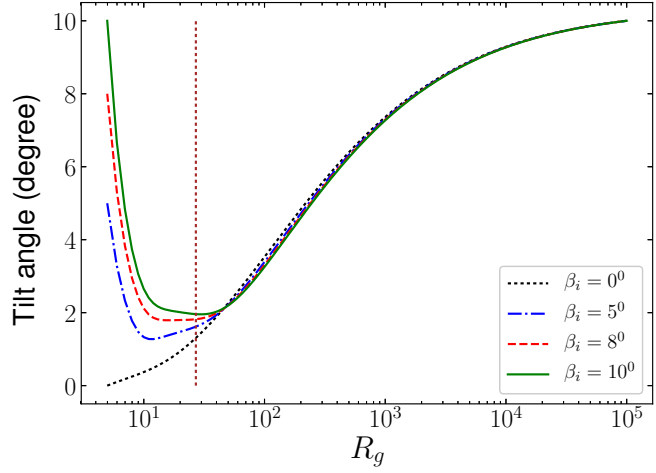}
\caption{Radial profiles of the disk tilt angle for different values of the disk inner edge tilt angle $\beta_i$. The other parameter values are $a=0.3$, $M=10M_{\odot}$, $\nu_2=10^{15}$ $\rm cm^2$ $s^{-1}$, $n=0.25$ and $z_{\rm in}=0.75$. The vertical line refers to the warp radius as estimated by the relation $\omega_p/\nu_2$. This figure exhibits the dip feature, and also there is a nonzero tilt at the inner disk for all the values of $\beta_i$ (see section \ref{res3} for details).}
\label{fig7}
\end{center}
\end{figure}
But this tilt angle also depends on $\nu_2$,
since a higher $\nu_2$ value would imply a larger viscous torque, which would alleviate the effect of the stronger LT torque 
(see Figure \ref{fig7a}). For example, when $\nu_2=10^{14}$ $\rm cm^2$ $s^{-1}$ (see Figure \ref{fig5}), we find that the alignment of the disk with the black hole equatorial plane occurs for $a\gtrsim0.3$, whereas for $\nu_2=10^{15}$ $\rm cm^2$ $s^{-1}$ (see Figure \ref{fig6}) this alignment occurs for $a\gtrsim0.8$. These show that the competition between LT torque and viscous torque determines the alignment of the disk.
\begin{figure}[h!]
\begin{center}
\includegraphics[width=0.49\textwidth]{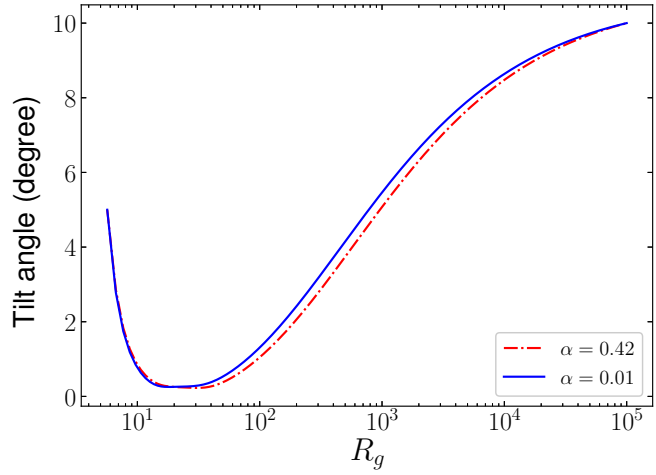}
\caption{Radial profiles of the disk tilt angle for different values of $\alpha$ or $n$. The other parameter values are $a=0.1$, $M=10M_{\odot}$, $\nu_2=10^{14}$ $\rm cm^2$ $s^{-1}$, $z_{\rm in}=0.3$, and $\beta_i=5^{\circ}$ (see section \ref{res3} for details).}
\label{fig8}
\end{center}
\end{figure}

Suppose the inner part of the disk is partially aligned for a set of parameter values. In such a case, as the above discussion indicates, the radius $R_{\rm align}$,
up to which the disk remains aligned, is higher for a higher value of the Kerr parameter. Note that, in our computation, we use 
$R_{\rm align}$ as the radius up to which the tilt angle is less than 0.01 degree
(implying an alignment for practical purposes).  
On the other hand, the characteristic warp radius $R_{\rm warp}$, 
inside which the LT effect dominates, is defined as the distance at which the timescale for warp diffusion (i.e., $R^2/\nu_2$) equals the local LT precession timescale (i.e., $R^3/\omega_p$) \citep{SF96,Lodato}. This gives
\begin{equation}\label{warping}
R_{\rm warp}=\frac{\omega_p}{\nu_2}.
\end{equation}
The alignment radius was approximated earlier by the warp radius
in various works \citep[e.g., ][]{Natarajan,Martin}, and the expression for $R_{\rm warp}$ was used in \cite{Martin} to calculate the alignment radius for the X-ray binary V4641 Sgr. But, in reality, the disk can remain significantly tilted at $R_{\rm warp}$. 
For example, according to the solution of \cite{SF96}, the tilt angle at this position is $\sim0.13$ times the tilt angle at the disk outer edge, although their solution may not 
be valid at the inner disk.
The relation between $R_{\rm align}$ and $R_{\rm warp}$ was found to be
$R_{\rm align}=0.165R_{\rm warp}$ by \cite{Armitage}, from comparing their numerical computations 
with the Equation~(\ref{warping}) (although \cite{Armitage} did not use the term $R_{\rm warp}$ explicitly). 
Note that \cite{Armitage} defined $R_{\rm align}$ as the radius up to which the tilt angle is small, 
but the value was not given.
We find this proportionality factor to be roughly $0.094$ from our computations.

In this context of disk alignment, here we qualitatively compare our results with those of \cite{Lodato}. 
Lodato and Pringle (2006) discussed their results in two regimes when $R_{\rm warp}>>R_{\rm in}$ (i.e., large warp radius case) and $R_{\rm warp}\simeq R_{\rm in}$ (i.e., small warp radius case). When the warp radius is large and $\nu_2/\nu_1$ is smaller, they 
found their tilt angle radial profile to deviate more from the same derived by \cite{SF96}. This is consistent with our work in the 
following way. The term associated with $C_1$ in the equation~(\ref{warp}), which was ignored in \cite{SF96},
is proportional to $(n+1)$ (see the term associated with $C_1$ in the equation (\ref{noeq})). Hence, for a higher value of 
$n$, implying a lower value of $\nu_2/\nu_1$, the deviation from the results obtained in \cite{SF96} is expected to be higher. 
When the warp radius is small, \cite{Lodato} found from their numerical investigations that 
the inner disk could be tilted with respect to the black hole spin. 
This result is similar to what we found, as smallness of warp radius implies a weak LT torque (for a fixed $\nu_2$; see the equation (\ref{warping})) or a strong viscous torque (for a fixed $a$), both would imply an inner disk tilt, as discussed above. 
The scaling relation mentioned above, i.e., $R_{\rm align}=0.094R_{\rm warp}$, is also in line with this.

Now we explore how the radial profile of the disk tilt angle is affected due to different choices of inner edge tilt angles, and $\alpha$ parameters (or $n$) of the disk. 
In the right panels of the Figures \ref{fig3} and \ref{fig4}, which are for nonzero disk inner edge
tilt angle ($\beta_i$), 
one finds a dip in the radial profile for the disk tilt angle in the analytical as well as numerical results. 
We explore the $\beta_i$ dependence of the plots more closely in Figure \ref{fig7}. 
We find that although the depth of the dip (i.e., the difference between $\beta_i$ and tilt angle at the dip) takes a higher value for higher $\beta_i$, the disk tilt angle profiles assume similar shapes far from the black hole. 
We note that, as $\beta_i$ increases, the dominance of the LT torque near the disk inner edge also increases.
This is because the strength of the LT torque depends also on the relative mismatch between the black hole spin axis 
and the disk angular momentum vector (see the equation (\ref{LT})). 
Hence, for a higher value of $\beta_i$, the LT torque is stronger near the inner edge, 
and it sharply reduces the mismatch in the disk orientation. The disk tilt angle attains
a minimum value, which may not be zero, depending on the magnitudes of $\nu_2$, $a$ and $M$.
However, the LT torque becomes weak far from the black hole, and the viscous torque plays the main role 
in deciding the radial tilt profile, resulting in similar profiles for different values of $\beta_i$.  

In Figure \ref{fig8}, we study the variation of the tilt angle profile for two different values of $n$ or $\alpha$. 
We find that the warp is slightly shifted towards the black hole for a higher value of $n$ (a lower value of $\alpha$ 
or a higher value of $\nu_1$ as $\nu_2$ is fixed), when the other parameters are fixed. We also find that 
the parameter $n$ or $\alpha$ (or $\nu_1$) affects the radial profile of the disk tilt angle weakly. 

\section{Summary and conclusions}\label{sum}
In this paper, we analytically solve the prograde warped accretion disk equation in the viscous regime for a slowly spinning Kerr black hole, and obtain an analytical expression for disk tilt angle up to the first order in Kerr parameter $a$. We take into account the contribution from the inner disk, which has been ignored in most earlier works. We also solve the warped disk equation numerically, and examine to what extent the analytical results, within their regime of validity, capture the essential features of this scenario by comparing the same with numerical results. We finally analyze the behavior of the radial profile of the disk tilt angle as a function of several parameters using our numerical results. The alignment of the inner disk strongly depends upon the relative dominance between LT torque (controlled mainly by $a$ and $M$), and viscous torque in the plane of the disk (controlled by $\nu_2$). We find that the inner disk could even be entirely misaligned for a reasonable range of parameter values, and report an empirical relationship between the alignment radius and the warp radius, which could be used to confront observations. There exists a critical value of Kerr parameter $a$, mainly 
depending upon the values of $\nu_2$ and $M$, beyond which the disk starts aligning itself with the 
black hole spin direction, i.e., the BP effect switches on. 

The inner accretion disk can play an important role to probe the physics of the strong gravity regime. 
A tilt in the inner disk with respect to the black hole spin axis can affect the spectral and timing 
properties of the X-ray emission through the LT precession, and hence can be particularly useful to study this regime.
A strong connection between the spectral (Fe K$\alpha$ line) and timing (quasi-periodic oscillation or QPO) 
features of X-ray emission has been reported in \cite{Miller}, in which they showed that the flux of the 
spectral Fe K$\alpha$ emission line from the Galactic black hole GRS 1915+105 varies
with the phase of low frequency QPOs in X-rays. Their results indicate that both the 
observed QPO and Fe K$\alpha$ line could originate due to LT precession of the warped inner accretion disk. 
Later, \cite{Schnittman} developed a simple model based on an inclined ring of hot gas orbiting around 
a Kerr black hole to explain the above mentioned connection, and used their model to make 
predictions on black hole parameters, spectral features, and light curves of similar X-ray binaries. 
Recently, \cite{Ingram} have also reported that the broad relativistic Fe line centroid energy from the 
accreting black hole H1743--322 systematically varies with the phase of a QPO. This could also be 
explained assuming LT precession of a tilted thin inner accretion disk, which being a reflector, can 
give rise to the observed Fe line (e.g., see sections 6.1 and 6.3 of \cite{Tom}). 
Therefore, as our results show that the inner disk may remain tilted for a reasonable range of parameter values, 
our solution of the radial profile of the disk tilt angle could be useful to observationally probe the 
strong gravity region around black holes.\\

C. C. gratefully acknowledges support from the National Natural Science Foundation of
China (NSFC), Grant No. 11750110410.

\end{document}